\documentclass[12pt]{article}

\usepackage{natbib,graphicx,setspace,lscape,longtable,epstopdf,xr}
\usepackage{mathrsfs,amsmath,amsthm,amssymb,color}
\usepackage{natbib,epsfig,graphicx,pdfpages}
\usepackage{rotating}
\usepackage{bm}
\usepackage{CJK}
\usepackage{ragged2e}
\usepackage{threeparttable}
\usepackage{multirow}
\usepackage{url}
\usepackage{natbib,graphicx,setspace,lscape,longtable,epstopdf,xr,rotating,epsfig,pdfpages,bm,ragged2e,geometry}
\usepackage{threeparttable,booktabs,multirow}
\usepackage[font=small,labelfont=bf]{caption}
\usepackage[noend, linesnumbered, ruled]{algorithm2e}
\usepackage{subfig}
\usepackage{hyperref}
\usepackage{booktabs}
\usepackage{makecell} 
\usepackage{arydshln}
\newlength\myindent
\allowdisplaybreaks[3] 
\usepackage{tikz}
\usepackage{bbding}

\usepackage{caption}
\usepackage{subfloat}
\usetikzlibrary{arrows,shapes,chains}
\usepackage{rotating}

\bibpunct{(}{)}{;}{a}{,}{,}

\newtheorem{theorem}{Theorem}

\newtheorem{definition}{Definition}

\newtheorem{corollary}{Corollary}
\newcommand{\csection}[1]
{\begin{center}
		\stepcounter{section}
		{\bf\large\arabic{section}. #1}
	\end{center}
}

\newtheorem{remark}{Remark}

\renewcommand{\arraystretch}{1.3}

\numberwithin{equation}{section}

\begin{document}
	\begin{center}
		{\bf\Large Data Privatization in Vertical Federated Learning with Client-wise Missing Problem}
	\end{center}
	
	\begin{center}
		Huiyun Tang$^{2}$, Long Feng$^{3}$, Yang Li$^{1,2}$, and Feifei Wang$^{1,2}$\footnote{Corresponding author: feifei.wang@ruc.edu.cn}\\
		{\it\small
			$^1$ Center for Applied Statistics, Renmin University of China, Beijing, China;\\
			$^2$ School of Statistics, Renmin University of China, Beijing, China;\\
			$^3$ Department of Statistics \& Actuarial Science, The University of Hong Kong, Hong Kong, China.
		}
	\end{center}
	
	\begin{singlespace}
		
\begin{abstract}
	Vertical Federated Learning (VFL) often suffers from client-wise missingness, where entire feature blocks from some clients are unobserved, and conventional approaches are vulnerable to privacy leakage. 
	We propose a Gaussian copula-based framework for VFL data privatization under missingness constraints, which requires no prior specification of downstream analysis tasks and imposes no restriction on the number of analyses. 
	To privately estimate copula parameters, we introduce a debiased randomized response mechanism for correlation matrix estimation from perturbed ranks, together with a nonparametric privatized marginal estimation that yields consistent CDFs even under MAR. 
	The proposed methods comprise VCDS for MCAR data, EVCDS for MAR data, and IEVCDS, which iteratively refines copula parameters to mitigate MAR-induced bias. Notably, EVCDS and IEVCDS also apply under MCAR, and the framework accommodates mixed data types, including discrete variables. 
	Theoretically, we introduce the notion of Vertical Distributed Attribute Differential Privacy (VDADP), tailored to the VFL setting, establish corresponding privacy and utility guarantees, and investigate the utility of privatized data for GLM coefficient estimation and variable selection. 
	We further establish asymptotic properties including estimation and variable selection consistency for VFL-GLMs. 
	Extensive simulations and a real-data application demonstrate the effectiveness of the proposed framework.
\end{abstract}


\end{singlespace}

\newpage

\csection{INTRODUCTION}

As vast quantities of sensitive data are collected and maintained in a distributed fashion, extracting accurate statistical insights while mitigating the risk of privacy breaches has become a paramount concern. Federated learning (FL) \citep{kairouz2021advances} has emerged as a compelling paradigm that facilitates collaborative model training across distributed data sources while preserving privacy and data security. 
Our research focuses on Vertical Federated Learning (VFL), a specialized category of federated learning in which data holders share the same set of samples but possess non-overlapping subsets of covariates \citep{liu2024vertical}. Such setting appears in many real applications, where the information describing an individual is collected and held by different entities. For example, a person's medical records are sensitive personal information that can be held by several clinics. 
Additionally, in cross-silo collaborative scenarios, organizations or institutions such as banks, hospitals, and government agencies, each holding small and fragmented datasets, are eager to seek data partners to jointly develop statistical or machine learning models, thereby enhancing data utility.
In VFL, since the participating entities are often reliable organizations which have access to substantial computational resources and high-bandwidth communication infrastructure, computational and communication constraints are generally not the primary bottlenecks in these settings.


VFL employs a fundamentally different training architecture compared to the classical FL framework, where data is partitioned by example. In VFL, it is typically assumed that the response variable is held by one participant, and this particular participant serves as the server which communicates with all the other participants (commonly referred to as \textit{clients}) to orchestrate the training procedure.
Depending on the specific training algorithm, clients exchange intermediate results rather than model parameters to facilitate model training across different clients. The choice of training algorithm is typically dictated by the underlying machine learning objective. Existing approaches include decision trees \citep{fang2021large}, random forests \citep{liu2020federated}, linear regression and K-means clustering \citep{huang2022coresets}, the Cox proportional hazards model \citep{dai2020verticox}, quantile regression \citep{fan2023residual}, and neural networks \citep{wang2023unified}.


In Vertical Federated Learning (VFL), although clients can collaboratively perform data analysis tasks without exposing their raw data, the training process requires frequent sharing and transmission of intermediate results, which poses privacy risks. In particular, \textit{honest-but-curious} participants may infer private information about other clients by observing the training process.
For instance, adversaries may reconstruct portions of the original data from transmitted gradient information \citep{zhu2019deep}, or infer whether specific data samples originate from a particular client based on intermediate results \citep{jin2021cafe, fu2022label}.
A common approach to mitigate privacy risks is incorporating Differential Privacy (DP) \citep{dwork2006differential} into the VFL framework. 
In general, DP ensures that the released statistical analysis results do not change much whether one particular record is present or absent in the underlying dataset, thereby preserving the sensitive information about the input data.
DP provides a rigorous mathematical formulation to quantify privacy loss and defines the probability bounds on privacy leakage in an algorithm. Due to its strong theoretical guarantees and practical feasibility, DP has become one of the most widely adopted privacy-preserving frameworks. It has also been deployed in real-world applications by major industry leaders \citep{ding2017collecting, Apple2017, Abowd2020}.


There are typically two streams of approaches about usage of DP in the context of VFL. The first involves designing algorithms that satisfy DP by introducing randomness or noise to obscure the influence of individual records on the output. For instance, \cite{hu2020learning} proposed a privacy-preserving ADMM algorithm for empirical risk minimization in VFL, where each client's transmitted values are perturbed with carefully calibrated Gaussian noise to ensure DP. Similarly, \cite{wang2020hybrid} developed a DP-based privacy-preserving stochastic gradient descent (SGD) algorithm for learning generalized linear models (GLMs) from vertically partitioned data while maintaining participant data confidentiality.
Despite their simplicity, these methods often suffer from degraded parameter estimation accuracy due to the added noise. 

The second major approach involves generating privacy-preserving synthetic data. Extensive research has been conducted on differentially private data release in centralized settings, where all data reside on a single machine \cite{zhang2024differentially, yang2024tabular}. However, synthesizing vertical data presents greater challenges, primarily due to the difficulty of capturing dependencies between covariates across datasets held by different clients.

Recently, studies on differentially private data construction under VFL have begun to emerge, which can be broadly classified into two categories. The first category is based on statistical modeling. \cite{mohammed2013secure} introduced an early method for releasing vertically partitioned data, utilizing a predefined attribute taxonomy tree as publicly available information. However, this approach is limited to two-party scenarios due to constraints imposed by the underlying cryptographic primitives. \cite{tang2019differentially} extended this method by employing a latent tree model to approximate the joint distribution of the integrated dataset, allowing it to handle a larger number of attributes. Nevertheless, its applicability remains constrained, as it is designed specifically for datasets with binary attributes and incurs significant communication and computational overhead due to the complexity of the cryptographic protocol. More recently, \cite{zhao2024vertimrf} proposed an approach based on a differentially private Markov Random Field (MRF), where a central server reconstructs a global MRF to capture correlations using locally shared MRFs and DP-protected sketches of client data. However, this method remains constrained to datasets with discrete attributes, limiting its broader applicability.

The second category leverages distributed generative adversarial networks (GANs) for collaborative synthetic data generation in a vertical setting \cite{jiang2023distributed, zhao2023gtv}. However, these GAN-based approaches struggle to capture dependencies among covariates across different clients. This limitation arises because the local discriminators are trained independently on each client's data, distinguishing between real and synthetic local samples without direct interaction across datasets.

In addition to the limitations discussed above, a critical yet underexplored challenge in VFL is client-wise missing data, a prevalent issue in real-world applications. This occurs when certain clients entirely lack variable information for some samples due to practical constraints in data collection.
In VFL literature, samples with fully observed covariates across all clients are referred to as aligned samples, and conventional VFL methods typically assume their availability. However, in practice, as the number of clients increases, it becomes more common for certain sources of information to be completely missing for a subset of samples, leading to a large proportion of non-aligned samples.

A straightforward approach to handling this issue is to discard non-aligned samples, but doing so results in a significant loss of information and can introduce bias in parameter estimation, ultimately degrading model performance. Thus, integrating all available data while maintaining privacy constraints is crucial for improving VFL effectiveness. Existing approaches in VFL literature primarily rely on deep learning and transfer learning techniques to infer missing features or their representations \citep{yang2022multi, kang2022fedcvt, ren2022improving, he2024hybrid}. While these methods can enhance predictive accuracy, they generally lack model interpretability, and they did not consider data privacy protection.

A similar concept in statistical literature is block-wise missing data, where certain blocks of features are unavailable when integrating multi-source datasets. Several approaches have been proposed to address this issue, including multiple block-wise imputation \citep{xue2021integrating}, GAN-based synthesis \citep{fang2024fragmgan}, semi-supervised learning \citep{song2024semi, li2024adaptive}, and projected estimating equations \citep{xue2025statistical}. However, these methods are designed for centralized settings, where all data is aggregated on a single server, and thus are not directly applicable to VFL, where privacy constraints prevent raw data sharing among clients.

Given the unique challenges posed by client-wise missing data, there is a pressing need for principled solutions specifically tailored to VFL. An ideal approach should effectively integrate fragmented data from multiple sources while balancing privacy preservation, interpretability, and computational efficiency, ensuring both theoretical soundness and practical feasibility in real-world applications.

\csection{METHODOLOGY}\label{sec2}
We propose a novel privacy-preserving framework tailored for the vertical federated learning (VFL) paradigm, built upon the Gaussian copula model. 
At a high level, assuming the variables in the original dataset follow a Gaussian copula model, synthetic samples can be generated by first drawing new samples from the Gaussian copula distribution and then mapping them to the original variables through the established correspondence between the latent Gaussian variables and the original variables.

\subsection{Problem Setup and Notation}

We consider a vertical federated learning framework comprising $K$ clients. Each client ($1 \leq k \leq K$) possesses a distinct subset of covariates, denoted as $\mathbf{X}_k = \{\mathbf{x}_{ki}\}_{i=1}^{N}$, for a shared cohort of $N$ subjects. Specifically, for each subject $i$, the covariate vector $\mathbf{x}_{ki} \in \mathbb{R}^{p_k}$ represents the $p_k$-dimensional feature set owned by client $k$. The complete covariate profile of subject $i$ is then given by  
$\mathbf{x}_i \triangleq (\mathbf{x}_i^{1\top}, \ldots, \mathbf{x}_i^{K\top})^{\top} \in \mathbb{R}^{p}$, 
where $p = \sum_{k=1}^{K} p_k$. Consequently, the aggregated covariate matrix, constructed by concatenating all clients' feature sets, is expressed as  
$\mathbf{X} = (\mathbf{X}_1,\dots, \mathbf{X}_K) \in \mathbb{R}^{N \times p}$.  
Without loss of generality, we assume that the response vector $\mathbf{y} \in \mathbb{R}^{N}$ is exclusively held by one of the $K$ clients, which consequently assumes the role of a central server, orchestrating the overall model training process.

In addition, suppose the response variable is fully observed by the central server for all subjects, while the covariates may be partially unavailable at the client level. Specifically, for a given subject, the covariate subset held by a particular client may be entirely missing. 
Define the missing indicator vector for each subject $i$ as $M_i = (M_{i1}, \ldots, M_{iK})^{\top}$, where $M_{ik} = 1$ if the $p_k$-dimensional covariates from the $k$-th client are missing and $M_{ik} = 0$ otherwise. The overall missing data pattern across all subjects is then represented by the matrix $M = (M_1, \ldots, M_N)^{\top}$.  
Further define $\Delta_{i,\text{obs}}$ and $\Delta_{i, \text{mis}}$ as the sets of clients for which the corresponding covariates are observed and missing, respectively, for subject $i$. Formally, $\Delta_{i,\text{obs}} = \{1\leq k \leq K: M_{ik} = 0\}, \quad \Delta_{i,\text{mis}} = \{1\leq k \leq K: M_{ik} = 1\}$. 
Accordingly, the covariate vector for subject $i$ can be rewritten as  
$\mathbf{x}_i = (\mathbf{x}_{i, \text{obs}}, \mathbf{x}_{i,\text{mis}})$,  
where $\mathbf{x}_{i,\text{obs}} = (\mathbf{x}_i^{k})_{k \in \Delta_{i,\text{obs}}}$ represents the observed covariates, and $\mathbf{x}_{i,\text{mis}} = (\mathbf{x}_i^{k})_{k \in \Delta_{i,\text{mis}}}$ corresponds to the missing ones. Notably, the dimensions of $\mathbf{x}_{i,\text{obs}}$ and $\mathbf{x}_{i,\text{mis}}$ generally vary across subjects.  

Our primary objective is to generate privatized pseudo-complete data in the presence of distributed multi-source covariates with client-specific missing observations. To this end, we develop a synthetic data generation framework that ensures data privacy using a copula-based model. Specifically, we first provide preliminaries on the techniques employed for data privatization in Section X. We then present the proposed algorithms tailored for MCAR and MAR missing mechanisms in Sections X and X, respectively.

%

\subsection{Preliminary}

Let the original dataset be denoted by $\mathbf{X} = (X_0, X_1, \ldots, X_p)$, where $X_0$ represents the response variable $Y$. 
The covariates may be of mixed types, including continuous, binomial, categorical, ordinal, and count variables. 
The procedure for handling discrete variables is provided in the Supplementary Material. 
The Gaussian copula models the $(p+1)$-dimensional vector as
\begin{align}\label{Eq3}
	\mathbf{Z} \sim \mathcal{N}_{p+1}(\mathbf{0}, \bm{\Omega}), \quad
	X_j = F_j^{-1}\!\left(\Phi(Z_j)\right), \quad j = 0, 1, \ldots, p.
\end{align}

Under this construction, the joint dependence structure among all variables across clients is captured via a Gaussian copula. This framework links the set of univariate marginals $\{F_j\}_{j=0}^p$ associated with each component of $\mathbf{X}$ to a multivariate distribution over latent Gaussian variables $\mathbf{Z}$, governed by a correlation matrix $\bm{\Omega}$. 
In this representation, each marginal distribution $F_j$ characterizes the distributional properties of $X_j$, while the correlation matrix $\bm{\Omega}$ encodes the global dependency structure among variables. When the marginals $X_j$ are normally distributed, $\bm{\Omega}$ coincides with the Pearson correlation matrix of $\mathbf{X}$; otherwise, it captures rank-based dependencies. Specifically, for $\bm{\Omega} = (r_{j_1 j_2})_{(p+1) \times (p+1)}$, the pairwise rank-based correlations-Kendall's $\tau$ and Spearman's $\rho$ are given by
\[
\tau_{j_1 j_2} = \frac{2}{\pi} \arcsin(r_{j_1 j_2}), 
\quad 
\rho_{j_1 j_2} = \frac{6}{\pi} \arcsin\!\left(\frac{r_{j_1 j_2}}{2}\right), \quad \forall j_1, j_2 \in \{0, \ldots, p\}~ \text{and}~ j_1 \neq j_2.
\]

Apparently, according to \eqref{Eq3}, given new latent Gaussian samples $\mathbf{z}_i \sim \mathcal{N}_{p+1}(\mathbf{0}, \bm{\Omega})$, the synthetic data can be obtained via the transformation $\tilde{x}_{ij} = F^{-1}_j(\Phi(z_{ij}))$ for each $j = 0,1,\cdots,p$.
Owing to their flexibility in modeling joint distributions with heterogeneous marginals, copula-based methods have been widely adopted for synthetic data generation in centralized settings, where multi-source datasets can be aggregated \citep{li2014differentially, gambs2021growing, jiang2023privacy}. However, these approaches are not directly applicable in the VFL setting due to the decentralized nature of the data and the lack of centralized access to the full joint distribution.

Copula-based modeling entails two primary components: estimation of the correlation structure $\bm{\Omega}$ and the marginal distributions $\{F_j\}_{j=0}^p$. The correlation matrix $\bm{\Omega} = (r_{ij})_{(p+1) \times (p+1)}$ is typically inferred via rank-based methods, with each entry estimated by the relationship $r_{ij} = \sin( \frac{\pi}{2} \tau_{ij})$, where $\tau_{ij}$ denotes the Kendall’s tau rank correlation coefficient between variables $X_i$ and $X_j$. In the context of VFL, the global correlation matrix $\bm{\Omega}$ must be collaboratively estimated across clients. To this end, each client shares perturbed ranks of its local variables with the server, enabling secure computation of cross-party rank correlations while enhancing privacy protection through randomized rank perturbation.

	\paragraph{Privatization of the Rank.}
To prevent potential privacy leakage from rank information, we apply a local differential privacy (LDP) mechanism to the ranks of each variable independently. For each variable $j$, its rank vector $R_j$ is encoded as a binary vector $V_j \in \{0,1\}^{N(N-1)/2}$, where each entry $v_{ii'}$ represents the outcome of a pairwise comparison between samples $i$ and $i'$:
if $X_{ij} > X_{i'j}$ for $1 \leq i < i' \leq N$, then $v_{ii'} = 1$; otherwise, $v_{ii'} = 0$.
The rank of $X_{ij}$ can be recovered via $R(X_{ij}) = 1 + \sum_{i' \neq i} c_{ii'}$, where $c_{ii'} = v_{ii'}$ if $i < i'$ and $c_{ii'} = 1- v_{i'i}$ if $i > i'$.

To protect these pairwise comparisons, we adopt the randomized response (RR) mechanism \citep{warner1965randomized, xu2025rate}, which perturbs each $v_{ii'}$ independently:
\begin{equation}\label{Eq1}
	\tilde{v}_{ii'} = 
	\begin{cases}
		v_{ii'}, & \text{with probability } 1 - \theta, \\
		1 - v_{ii'}, & \text{with probability } \theta,
	\end{cases}
\end{equation}
where $\theta = \left(1 + \exp(\epsilon)\right)^{-1}$, and $\epsilon$ is the local privacy budget. For any $\epsilon > 0$, we have $\theta \in (0, 1/2)$.
The privatized rank is then $\widetilde{R}(X_{ij}) = 1 + \sum_{i' \neq i} \tilde{c}_{ii'}$, with $\tilde{c}_{ii'} = \tilde{v}_{ii'}$ if $i < i'$ and $\tilde{c}_{ii'} = 1- \tilde{v}_{i'i}$ if $i > i'$.

The RR mechanism satisfies $\epsilon$-LDP given $\theta = \left(1 + \exp(\epsilon)\right)^{-1}$ \citep{yang2024local, wang2017locally}, guaranteeing privacy protection for all observed pairwise comparisons. Due to its effectiveness, the RR mechanism has been widely adopted in practice, including in Google Chrome browser \citep{erlingsson2014rappor},  macOS \citep{tang2017privacy}, and recommender system \citep{kalloori2018eliciting}. 
Prior work \citep{xu2025rate} also shows that the RR mechanism outperforms additive noise mechanisms such as the Laplace mechanism \citep{dwork2006differential} in preserving the relative order of entries in the original rankings, even when achieving the same privacy budget.

Denote by $\widehat{\bm{\Omega}}$ the empirical rank correlation matrix computed from the true (unperturbed) ranks, and by $\widetilde{\bm{\Omega}}$ the privatized estimator constructed from perturbed ranks. Theorem \ref{theorem1} establishes the convergence rate of the $l_{\infty}$-norm error between $\widetilde{\bm{\Omega}}$ and $\widehat{\bm{\Omega}}$. According to \eqref{eq:first}, the convergence rate of the relative error of $\widetilde{\bm{\Omega}}$ is a constant given any fix privacy parameter $\epsilon_{1}$, implying that $\widetilde{\bm{\Omega}}$ does not converge to $\widehat{\bm{\Omega}}$ even as $N \rightarrow \infty$. This inconsistency arises because the RR mechanism shifts the distribution of pairwise comparisons, introducing systematic bias into subsequent parameter estimation.
 To address the issue, we propose a \emph{debiased} RR mechanism. After obtaining the perturbed pairwise comparison vector 
  $\widetilde{V}_j$ via \eqref{Eq1}, we apply the following correction step:
\begin{equation}\label{Eq2}
	\tilde{v}^d_{ii'} = \frac{\tilde{v}_{ii'} - \theta}{1- 2 \theta}, \quad \text{for}~ i < i'.
\end{equation}
This adjustment yields an unbiased estimator for the original pairwise comparisons $v_{ii'}$. Denote by $\widetilde{R}_j^d$ the rank vector derived from the debiased $\widetilde{V}^d_j$. Then $\mathbb{E}(\widetilde{R}^d_j) = R_j$, where the expectation is taken with respect to the randomness of the privacy mechanism. As shown in \eqref{eq:second} of Theorem \ref{theorem1}, the resulting debiased estimator $\widetilde{\bm{\Omega}}^d$ is consistent and can asymptotically recover $\widehat{\bm{\Omega}}$ given sufficiently large $N$, overcoming the convergence barrier of the classical RR mechanism.

Regarding the estimation of marginal distributions, a common practice is to fix the marginals at their empirical distribution functions (ECDFs), denoted by $\{\widehat{F}_j\}_{j=0}^p$. According to classical results such as the Glivenko-Cantelli Theorem, $\{\widehat{F}_j\}_{j=0}^p$ constitutes a consistent estimator of the true marginal distributions $\{F_j\}_{j=0}^p$ when the dataset is fully observed or when the missing data mechanism satisfies the missing completely at random (MCAR) assumption \citep{little2019statistical}. 
We then employ the ECDF $\widehat{F}_j$ to generate privatized data. Since $\widehat{F}_j$ is a step function, it lacks a well-defined continuous inverse. A common remedy is to fit a monotone interpolating spline to the set $\{(X_{ij}, \widehat{F}_j(X_{ij}))\}_{i=1}^N$, and, for notational simplicity, we continue to denote the resulting smoothed function by $\widehat{F}_j$. 
However, $\widehat{F}_j$ is not differentially private. To avoid the leakage of sensitive information when applying the ECDFs $\{\widehat{F}_j\}_{j=0}^p$, we approximate each $\widehat{F}_j$ using privatized Bernstein polynomials \citep{phillips2003bernstein, alda2017bernstein}.

\paragraph{Privatization of the Marginal Distribution.} 
Specifically, we first rescale $(X_{1j},\cdots, X_{Nj})$ to the interval $[0,1]$, and then approximate $\widehat{F}_j(x)$ using the Bernstein operator:
\begin{equation*}
	B_{l}\left(\widehat{F}_j\right)(x):= \sum_{v=0}^{l}\widehat{F}_j\left(\frac{v}{l}\right)b_{v,l}(x), \quad b_{v,l}(x) = \binom{l}{v}x^v(1-x)^{l-v}, \quad v = 0,\cdots,l.
\end{equation*}
According to \cite{alda2017bernstein}, we can achieve differential privacy by adding noise to the Bernstein coefficients. That is,
\begin{equation*}
	\widetilde{B}_{l}\left(\widehat{F}_j\right)(x) = \sum_{v=0}^{l} \left(\widehat{F}_j\left(\frac{v}{l}\right) + \varepsilon_v\right)b_{v,l}(x),
\end{equation*}
where $\varepsilon_v \overset{i.i.d.}{\sim} \operatorname{Lap}(\Delta \widehat{F}_j \cdot (l+1)/\epsilon)$, and $\Delta \widehat{F}_j$ denotes the $\ell_1$-sensitivity, defined as
\begin{equation*}
	\Delta \widehat{F}_j := \sup_{X_j \sim X_j^{'}} \left|\widehat{F}_j(x) - \widehat{F}_j(x')\right|,
\end{equation*}
with $X_j \sim X_j^{'}$ indicating that $X_j$ and $X_j^{'}$ differ in at most one element belonging to a single person. Apparently, for ECDFs, we have $\Delta \widehat{F}_j = \frac{1}{N}$.

Furthermore, suppose $\widehat{F}_j$ is $(h,T)$-smooth, meaning it possesses continuous derivatives up to order $h$ (for some positive integer $h$), and all such derivatives are bounded in absolute value by $T$. In this case, a more refined approximation can be obtained using the Iterated Bernstein operator \citep{micchelli1973saturation}:
\begin{equation}\label{Eq9}
	\widetilde{B}_{l}^{(h)}\left(\widehat{F}_j\right)(x) := \sum_{v=0}^{l} \left(\widehat{F}_j\left(\frac{v}{l}\right) + \varepsilon_v\right) b^{(h)}_{v,l}(x), \quad b^{(h)}_{v,l}(x) = \sum_{i=1}^{h} \binom{h}{i}(-1)^{i-1} B_l^{i-1}(b_{v,l}; x),
\end{equation}
where $B_l^i = B_l \odot B_l^{i-1}$, and $B_l^0 = I$ denotes the identity operator.

We denote the final privatized estimator of the marginal CDF as $\widehat{F}_j^{\text{priv}} := \widetilde{B}_{l}^{(h)}\left(\widehat{F}_j\right)(x)$.
According to \cite{alda2017bernstein}, the estimator $\widehat{F}_j^{\text{priv}}$ satisfies $\epsilon$-differential privacy. In addition, Corollary \ref{corol-1} characterizes the overall error, which includes both the approximation error and the noise-induced error.

\subsection{The Vertical Copula-based Data Sanitization Method under MCAR}

\begin{figure}[htb]
	\centering
	\includegraphics[width=0.9\linewidth]{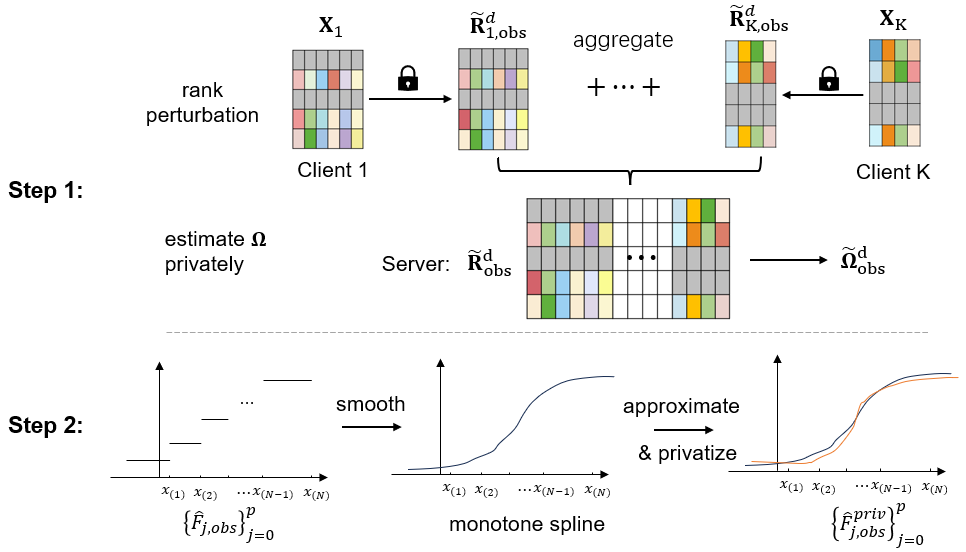}
	\caption{Illustration of the Vertical Copula-based Data Sanitization (VCDS) procedure under the MCAR mechanism, where gray cells denote missing entries.}
	\label{f1}
\end{figure}

Based on the above analysis, we propose the Vertical Copula-based Data Sanitization (VCDS) method to handle client-wise MCAR missingness. 
As illustrated in Figure \ref{f1}, the method consists of two main stages: (1) local rank perturbation and dependence estimation; (2) marginal distribution estimation and privatization.
The more detailed procedure is outlined in Algorithm \ref{alg1}. The process begins with each client $k$ perturbing the ranks of its local variables on observed entries using the debiased RR mechanism, yielding the perturbed observed rank matrix $\widetilde{\mathbf{R}}^{d}_{k,\text{obs}}$. In parallel, each client computes the ECDF $\widehat{F}_{j,\text{obs}}$ for each local variable employing only observed values, and applies monotone spline smoothing to ensure invertibility. To achieve differential privacy, the smoothed ECDF is approximated using privatized Bernstein polynomials, producing $\widehat{F}^{\mathrm{priv}}_{j,\text{obs}}(x) = \widetilde{B}^{(h)}_l(\widehat{F}_{j,\text{obs}})(x)$.
Next, the server collects the perturbed rank matrices from all clients and concatenates them into $\widetilde{\mathbf{R}}^d_{\text{obs}} = (\widetilde{\mathbf{R}}^d_{1,\text{obs}}, \dots, \widetilde{\mathbf{R}}^d_{K,\text{obs}})$. Based on these perturbed ranks, the server computes the rank correlation matrix $\widetilde{\bm{\Omega}}^d_{\text{obs}}$. Then, the server samples $\mathbf{Z}^{\mathrm{new}} \in \mathbb{R}^{N' \times (p+1)}$ from the multivariate normal distribution $\mathcal{N}_{p+1}(\mathbf{0}, \widetilde{\bm{\Omega}}^d_{\text{obs}})$ and sends the corresponding submatrix $\mathbf{Z}^{\mathrm{new}}_k$ to each client $k$. Upon receiving $\mathbf{Z}^{\mathrm{new}}_k$, each client applies the inverse of its privatized marginal distribution function to transform the Gaussian scores back to the original data domain: $\widetilde{X}_{j}^{\mathrm{priv}} = (\widehat{F}^{\text{priv}}_{j,\text{obs}})^{-1} \big( \Phi(Z^{\text{new}}_{j}) \big)$. This yields the privatized pseudo-data $\widetilde{\mathbf{X}}_k^{\mathrm{priv}}$ for each client. 

The VCDS algorithm ensures that sensitive raw data never leave the local clients. Ranks are perturbed under local differential privacy, marginal distributions are privatized via Bernstein polynomial approximation, and the dependence structure is reconstructed on the server using only perturbed statistics. Theorem \ref{thm1} shows that the generated synthetic data preserve both the marginal distributions and the cross-party dependence structure while satisfying formal privacy guarantees. Theorem \ref{thm2} provides a theoretical upper bound on the utility of the VCDS mechanism, quantifying how well the raw data distribution is preserved under privacy protection.

\begin{algorithm}[!h]
	\caption{The VCDS Algorithm}
	\label{alg1}
	\KwIn{Local raw observed data $\mathbf{X}_{k,\text{obs}}$ for $k=1,2,\dots,K$} and $Y$
	\KwOut{Privatized pseudo data $\widetilde{\mathbf{X}}_k^{\text{priv}}$ for $k=1,2,\dots,K$}	and $\widetilde{Y}^{\text{priv}}$
	\For{each client $k \in \{1,2,\dots,K\}$}{
		Perturb the ranks of each local variable using the debiased RR mechanism to obtain perturbed ranks $\widetilde{\mathbf{R}}^{d}_{k,\text{obs}}$, and send them to the server\;
		Compute the ECDF $\widehat{F}_{j,\text{obs}}$ for each local variable and apply monotone smoothing\;
		Approximate each $\widehat{F}_{j,\text{obs}}$ via privatized Bernstein polynomials to obtain: $\widehat{F}^{\text{priv}}_{j,\text{obs}}(x) = \widetilde{B}^{(h)}_l(\widehat{F}_{j,\text{obs}})(x)$.
	}
	\textbf{On the server side:} \\
	Aggregate the rank matrices as $\widetilde{\mathbf{R}}^d_{\text{obs}} := (\widetilde{\mathbf{R}}^d_{1,\text{obs}}, \dots, \widetilde{\mathbf{R}}^d_{K,\text{obs}})$\;
	Compute the copula correlation matrix estimator $\widetilde{\bm{\Omega}}^d_{\text{obs}}$ from $\widetilde{\mathbf{R}}^d_{\text{obs}}$\;
	Draw a new sample $\mathbf{Z}^{\text{new}} \in \mathbb{R}^{N' \times (p+1)}$ from $\mathcal{N}_{p+1}(\mathbf{0}, \widetilde{\bm{\Omega}}^d_{\text{obs}})$, and return the corresponding submatrices $\mathbf{Z}^{\text{new}}_k$ to client $k$ for $k=1,2,\dots,K$\;
	\For{each client $k \in \{1,2,\dots,K\}$}{
		Generate the privatized synthetic data for each variable as	$\widetilde{X}_{j}^{\text{priv}} = (\widehat{F}^{\text{priv}}_{j,\text{obs}})^{-1} \big( \Phi(Z^{\text{new}}_{j}) \big)$.
	}
\end{algorithm}

Note that Algorithm \ref{alg1} is built upon the ECDF, fixing each $F_j$ at its empirical cumulative distribution function. However, the ECDF can incur substantial bias when the missing data mechanism is not completely at random, such as under the missing at random (MAR) scenario. To address this limitation, we propose an extended version of the Vertical Copula-based Data Sanitization method, referred to as EVCDS. Unlike the original approach, EVCDS does not rely on the ECDFs. Instead, it introduces a novel nonparametric estimator for the marginal distributions, which yields computationally efficient and consistent estimates of $\{F_j\}_{j=0}^p$ even in the presence of MAR data.

\subsection{The Extended Vertical Copula-based Data Sanitization Method under MAR}

In terms of the estimation of marginal distributions under more complex missing data mechanisms, inspired by the ``margin adjustment" approach proposed by \citet{feldman2024nonparametric}, we estimate the marginal CDFs $\{F_j\}_{j=0}^p$ by utilizing perturbed ranks together with the latent variables $\mathbf{Z}$. This approach leverages the insight that, under model \eqref{Eq3}, each $Z_j$ is a non-decreasing transformation of the corresponding variable $X_j$. More precisely, for each $j$, there exists a strictly increasing function $G_j(\cdot)$, possibly unknown, such that $X_j = G_j(Z_j)$ for $j = 0, 1, \ldots, p$.
As a consequence, when both $\{X_{ij}\}_{i=1}^N$ and $\{Z_{ij}\}_{i=1}^N$ are sorted, the position of $\max\{Z_{ij}: X_{ij} \leq x\}$ within $\{Z_{ij}\}_{i=1}^N$ coincides with that of $\max\{X_{ij}: X_{ij} \leq x\}$ within $\{X_{ij}\}_{i=1}^N$ for any $x$ exceeding the minimum of $\{X_{ij}\}_{i=1}^N$. Define
\begin{equation}\label{Eq7}
	\widetilde{Z}_j(x) = \max\left\{ \{Z^{'}_{ij}: \widetilde{R}(X_{ij})\leq \widetilde{R}(x)\} \cup \{Z_{ij}^{'}: \widetilde{R}(X_{ij}) = \min_{1 \le i \le N} \widetilde{R}(X_{ij})\}, i=1,\cdots,N \right\},
\end{equation}
where $\widetilde{R}(\cdot)$ denotes the perturbed rank function. The privatized marginal distribution estimator is then given by $\widetilde{F}_j(x):= \Phi(\widetilde{Z}_j(x))$. In the presence of missing data, $\widetilde{Z}_j(x)$ is calculated as
{\small 
	\begin{equation}\label{Eq10}
		\widetilde{Z}_j(x) = \max\left\{ \{Z^{'\text{obs}}_{ij}: \widetilde{R}(X_{ij}^{\text{obs}})\leq \widetilde{R}(x)\} \cup \{Z_{ij}^{'\text{obs}}: \widetilde{R}(X_{ij}^{\text{obs}}) = \min_{1 \le i \le N} \widetilde{R}(X_{ij}^{\text{obs}})\}, i=1,\cdots,N \right\}.
	\end{equation}
}

According to Eq.~\eqref{Eq4} in Theorem~\ref{theorem2}, we have $\widetilde{F}_j(x) \overset{\mathbb{P}}{\rightarrow} F_j(x)$, implying that $\widetilde{F}_j(x)$ is a consistent estimator of the true CDF $F_j(x)$. Crucially, this consistency remains valid under MAR. For clarity, we present the result in Theorem~\ref{theorem2-1} for the case $p=2$, where one variable follows an MAR mechanism; the extension to higher dimensions is straightforward.

Nevertheless, $\widetilde{F}_j(x)$ may fail to be a valid distribution function because it is not guaranteed to be monotonic. In particular, for any $x_1 < x_2$, it is possible that $\widetilde{F}_j(x_1) > \widetilde{F}_j(x_2)$ with positive probability when the privacy parameter $\epsilon$ and the sample size $N$ are fixed.  To address this issue, we construct a perturbed version of the variable $X_j$, denoted by $\widetilde{X}_j$, by reordering the original sample $\{X_{ij}\}_{i=1}^N$ according to their perturbed ranks. Specifically, let $\mathbf{x} = \{X_{1j}, \dots, X_{Nj}\} = \{X_{(R(X_{1j}))}, \dots, X_{(R(X_{Nj}))}\}$, and define the perturbed sequence as  
\begin{equation}\label{Eq8}
	\widetilde{\mathbf{x}} := \{X_{(\widetilde{R}(X_{1j}))}, \dots, X_{(\widetilde{R}(X_{Nj}))}\}.
\end{equation}
Under this construction, the mapping $\widetilde{x} \mapsto \widetilde{F}_j(\widetilde{x})$ is guaranteed to be non-decreasing, as the ranks of the inputs are aligned with those used in computing their function values. Moreover, Eq.~\eqref{Eq5} in Theorem~\ref{theorem2} ensures that $\widetilde{F}_j(\widetilde{x})$ attains the same convergence rate as $\widetilde{F}_j(x)$.

Notably, although $\{\widetilde{X}_{ij}\}_{i=1}^N$ are no longer linked to their original indices, thereby protecting individual identities, the variable values themselves may still leak sensitive information.  To further enhance the privacy protection of the original values $\{X_{ij}\}_{i=1}^N$, we approximate $\widetilde{F}_j(\tilde{x})$ by means of the privatized Bernstein polynomials as described in Eq. \eqref{Eq9}. We denote the resulting privatized estimator of the marginal CDF as $\widetilde{F}_j^{\text{priv}} := \widetilde{B}_{l}^{(h)}\left(\widetilde{F}_j\right)(\tilde{x})$.
According to \cite{alda2017bernstein}, the estimator $\widetilde{F}_j^{\text{priv}}$ satisfies $\epsilon$-differential privacy and admits the error bound stated in Corollary \ref{corol-2}.

In practice, however, the function $G_j(\cdot)$ is unknown a priori, which prevents us from directly generating latent variables $Z_j$ that satisfy the desired rank-consistency condition. To remedy this, we first draw an initial sample $\mathbf{Z}$ from $\mathcal{N}_{p+1}(\mathbf{0}, \widetilde{\bm{\Omega}}^d)$, and then transform it to ensure the rank structure matches that of $\mathbf{X}$.

Note that $Z_j$ being a non-decreasing transformation of $X_j$ implies the equivalence of rank ordering: $X_{ij} < X_{i'j} \;\Leftrightarrow\; Z_{ij} < Z_{i'j}$.
Define the event
$$
\mathbb{D}(\mathbf{X}):=\left\{\mathbf{Z} \in \mathbb{R}^{N \times(p+1)}: \max _{\left\{i^{\prime}: X_{i^{\prime} j}<X_{i j}\right\}} Z_{i^{\prime} j}<Z_{i j}<\min _{\left\{i^{\prime}: X_{i j}<X_{i^{\prime} j}\right\}} Z_{i^{\prime} j}, \quad \forall i \neq i^{\prime}, 0 \leq j \leq p\right\} .
$$
When $\mathbf{Z} \in \mathbb{D}(\mathbf{X})$, the rank ordering of each variable in the original dataset $\mathbf{X}$ is preserved in $\mathbf{Z}$. Specifically, for each $Z_{ij}$, its value lies within the interval determined by the latent variables corresponding to the adjacent ranks of $X_{ij}$. Thus, any $\mathbf{Z}$ satisfying $\mathbf{Z} \in \mathbb{D}(\mathbf{X})$ is suitable for subsequent marginal distribution estimation.

Since the initial draw $\mathbf{Z} \sim \mathcal{N}_{p+1}(\mathbf{0}, \widetilde{\bm{\Omega}}^d)$ generally does not satisfy $\mathbf{Z} \in \mathbb{D}(\mathbf{X})$, a naive approach is to reorder $\mathbf{Z}$ dimension-wise according to the ranks of $\mathbf{X}$. However, it only preserves the marginal order of each variable, but destroys the dependence structure between variables.
 To overcome this, we perform a sample-wise adjustment on each $Z_j$ to obtain $\mathbf{Z}'$ such that $\mathbf{Z}' \in \mathbb{D}(\mathbf{X})$ while preserving as much of the original dependence structure as possible.
Specifically, for each $j$ and sample $i$, we determine the interval 
\begin{equation}\label{Eq6}
z_l = \max \left\{ Z_{i'j} : \widetilde{R}(X_{i'j}) < \widetilde{R}(X_{ij}),\; i' \neq i \right\}, \quad
z_u = \min \left\{ Z_{i'j} : \widetilde{R}(X_{i'j}) > \widetilde{R}(X_{ij}),\; i' \neq i \right\},
\end{equation}
where $\widetilde{R}(\cdot)$ is the perturbed rank. Then, we sample from the truncated conditional normal distribution to obtain
$Z'_{ij} \sim \mathcal{N}(\mu_{ij}, \sigma_j^2) \, \mathbb{I}(z_l, z_u)$,
where the conditional mean and variance are given by
\[
\mu_{ij} = \widetilde{\bm{\Omega}}^d_{j,-j} (\widetilde{\bm{\Omega}}^d)_{-j,-j}^{-1} \mathbf{Z}_{i,-j}, 
\quad
\sigma_j^2 = \widetilde{\bm{\Omega}}^d_{j,j} - \widetilde{\bm{\Omega}}^d_{j,-j} (\widetilde{\bm{\Omega}}^d)_{-j,-j}^{-1} \widetilde{\bm{\Omega}}^d_{-j,j}.
\]
This adjustment guarantees that the corrected latent matrix $\mathbf{Z}'$ preserves the rank structure of $\mathbf{X}$ while restoring and retaining as much of the original multivariate normal dependence structure as possible. In practice, this is implemented sequentially by first sorting $Z_1$ to match the rank of $X_1$, then sampling each subsequent $Z'_j$ conditionally on the previously adjusted variables $Z'_{1}, \ldots, Z'_{j-1}$. 
The adjusted latent variables $\mathbf{Z}^{'}$ are then combined with the perturbed ranks to construct $\widetilde{F}_j^{\text{priv}}$.

\paragraph{The EVCDS Algorithm.}

\begin{figure}[htb]
	\centering
	\includegraphics[width=0.9\linewidth]{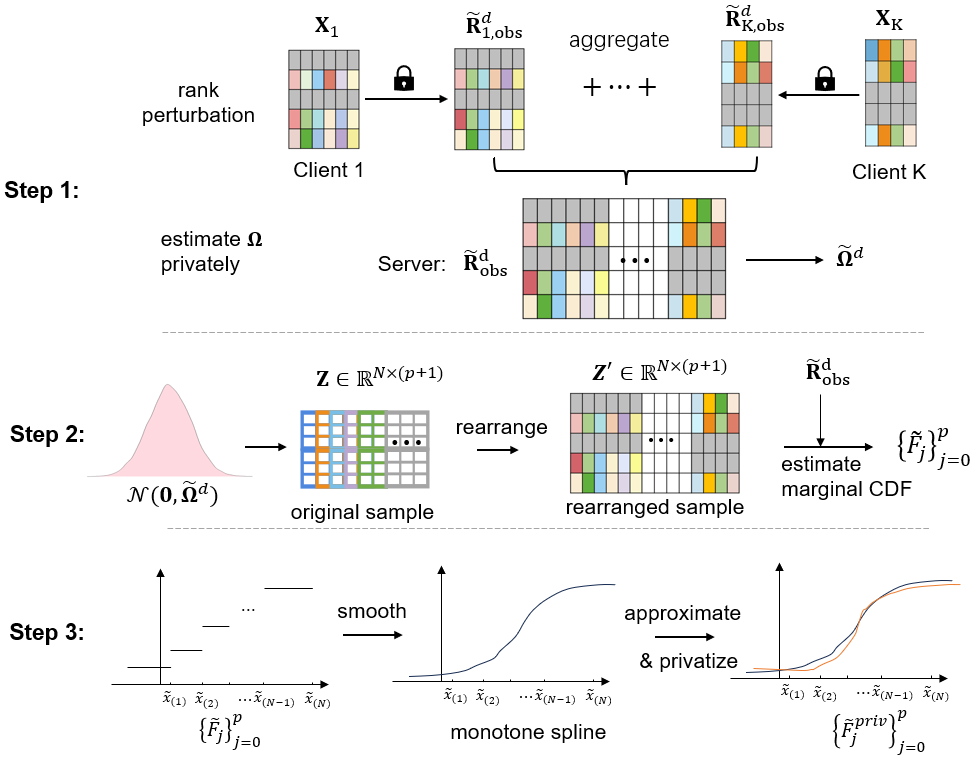}
	\caption{Illustration of the Extended Vertical Copula-based Data Sanitization (EVCDS) method under MAR, where gray cells represent missing entries.}
	\label{f2}
\end{figure}

As illustrated in Figure~\ref{f2}, the extended VCDS (EVCDS) method retains the core components of the original VCDS framework, including rank perturbation and correlation estimation in Step 1, and the privatization of estimated marginal distributions in Step 3. Compared with VCDS, EVCDS refines the procedure in Step 2 by avoiding independent marginal estimation for each variable. Instead, it explicitly incorporates correlation information into the nonparametric estimation of the cumulative distribution functions (CDFs). This is achieved by introducing Gaussian latent variables that share the same dependence structure as the original variables. The marginal distributions are then estimated jointly using these latent variables together with the perturbed ranks. Such a design can yield consistent estimators of $\{F_j\}_{j=0}^p$ even under MAR mechanisms.

The detailed procedure is summarized in Algorithm~\ref{alg2}. Following the same local client processing and server-side aggregation/correlation estimation steps as in Algorithm~\ref{alg1}, the server generates a Gaussian latent matrix $\mathbf{Z}^{'}$ that is necessary for marginal CDF estimation. Based on the estimated copula correlation $\widetilde{\bm{\Omega}}^d_{\text{obs}}$ obtained from the observed data (for notational simplicity, we omit the subscript $_{\text{obs}}$ and denote it as $\widetilde{\bm{\Omega}}^d$), the server draws an initial latent sample $\mathbf{Z} \sim \mathcal{N}_{p+1}(\mathbf{0}, \widetilde{\bm{\Omega}}^d)$. These latent variables are then reordered and sequentially adjusted to satisfy rank constraints derived from the privatized ranks. 
The adjustment is performed by sampling each element $Z_{ij}^{'}$ from a truncated conditional normal distribution with parameters computed based on previously adjusted variables, thereby ensuring consistency with the rank information while approximately preserving the original correlation structure. Then the adjusted latent matrix $\mathbf{Z}^{'}$ is partitioned and sent back to the respective clients for marginal distribution estimation. During synthetic data generation, unlike Algorithm \ref{alg1} where the newly drawn latent sample $\mathbf{Z}^{\text{new}}$ is directly fed into the inverse marginal CDFs, EVCDS applies the same adjustment procedure to $\mathbf{Z}^{\text{new}}$ to ensure rank consistency, resulting in higher-quality synthetic data. Note that the sample size of synthetic dataset produced by Algorithm \ref{alg2} can be a multiple of the original dataset size.

Corollary~\ref{coro3} establishes that the EVCDS method satisfies rigorous differential privacy guarantees. Notebaly, the additional procedure introduced to consistently estimate marginal distributions under the MAR mechanism does not incur any extra privacy cost. Hence, EVCDS shares the same privacy budget as the VCDS method.
Theorem~\ref{thm3} further characterizes the utility of EVCDS under client-wise MAR missingness by evaluating the KL divergence between the density fitted on the complete data (without missingness) and that fitted on the privatized data generated by Algorithm~\ref{alg2}. The theorem shows that EVCDS achieves asymptotic fidelity up to the MAR bias, with the discrepancy arising from the MAR bias induced by the estimation of the correlation matrix $\widetilde{\bm{\Omega}}^d$ based on the observed data. To mitigate this bias, we further propose the Iterated EVCDS Algorithm~\ref{alg3}, which iteratively refines the correlation estimation and thereby enhances the overall performance of EVCDS.

\begin{algorithm}[!h]
	\caption{The EVCDS Algorithm}
	\small
	\label{alg2}
	\KwIn{Local raw observed data $\mathbf{X}_{k,\text{obs}}$ for clients $k=1,2,\ldots,K$} and $Y$
	\KwOut{Privatized pseudo data $\widetilde{\mathbf{X}}_k^{\text{priv}}$ for clients $k=1,2,\ldots,K$ and $\widetilde{Y}^{\text{priv}}$}
	\For{each client $k \in \{1,\ldots,K\}$}{
		Perturb the ranks of each local variable to obtain $\widetilde{\mathbf{R}}_{k,\text{obs}}$, then apply the debiased RR mechanism to obtain $\widetilde{\mathbf{R}}_{k,\text{obs}}^d$\;
		Send $\widetilde{\mathbf{R}}_{k,\text{obs}}^d$ to the server\;
	}
	\textbf{Server side:} \\
	Aggregate rank matrices as $\widetilde{\mathbf{R}}^d_{\text{obs}} := (\widetilde{\mathbf{R}}_{1,\text{obs}}^d, \ldots, \widetilde{\mathbf{R}}_{K,\text{obs}}^d)$\;
	Compute the copula correlation matrix estimator $\widetilde{\bm{\Omega}}^d$ from $\widetilde{\mathbf{R}}^d_{\text{obs}}$\;
	Draw a sample $\mathbf{Z} \in \mathbb{R}^{N \times (p+1)}$ from $\mathcal{N}_{p+1}(\mathbf{0}, \widetilde{\bm{\Omega}}^d)$\;
	Sort $Z_0$ to match the order of $\widetilde{R}^d_{0,\text{obs}}$\;
	\For{$j = 1$ \KwTo $p$}{
		Sort $Z_j$ according to $\widetilde{R}^d_{j,\text{obs}}$\;
		\For{$i = 1$ \KwTo $N$}{
			Compute the truncated interval $(z_l, z_u)$ on the observed data as defined in Equation \eqref{Eq6}\;
			Compute conditional mean and variance:
			\[
			\mu_{ij} = \widetilde{\bm{\Omega}}^d_{j,1:(j-1)} \big(\widetilde{\bm{\Omega}}^d_{1:(j-1),1:(j-1)}\big)^{-1} \mathbf{z}^{(seq)}, \quad
			\sigma_j^2 = \widetilde{\bm{\Omega}}^d_{j,j} - \widetilde{\bm{\Omega}}^d_{j,1:(j-1)} \big(\widetilde{\bm{\Omega}}^d_{1:(j-1),1:(j-1)}\big)^{-1} \widetilde{\bm{\Omega}}^d_{1:(j-1),j},
			\]
			where $\mathbf{z}^{(seq)} = (Z'_{i1}, \ldots, Z'_{i(j-1)})$ and $Z'_{i0} = Z_{i0}$\;
			Sample $Z'_{ij} \sim \mathcal{N}(\mu_{ij}, \sigma_j^2)\mathbb{I}(z_l, z_u)$ \;
		}
	}
	Obtain the adjusted latent sample matrix $\mathbf{Z}'$, then distribute the corresponding submatrices $\mathbf{Z}'_k$ back to each client $k$\;
	\For{each client $k \in \{1,\ldots,K\}$}{
		Estimate marginal distributions by
		$\widetilde{F}_j(\tilde{x}) = \Phi\big(\widetilde{Z}_j(\tilde{x})\big)$,
		where $\widetilde{Z}_j(\cdot)$ and $\tilde{x}$ are defined in Equations \eqref{Eq10} and \eqref{Eq8}\;
		Approximate each $\widetilde{F}_j$ using privatized Bernstein polynomials:
		$\widetilde{F}_j^{\mathrm{priv}}(\tilde{x}) = \widetilde{B}_l^{(h)}(\widehat{F}_j)(\tilde{x})$.
	}	
	\textbf{Server side:} \\
	Draw a new sample $\mathbf{Z}^{\mathrm{new}} \sim \mathcal{N}_{p+1}(\mathbf{0}, \widetilde{\bm{\Omega}}^d)$,  
	rearrange $\mathbf{Z}^{\mathrm{new}}$ following steps similar to lines 8--14, and distribute submatrices $\mathbf{Z}^{\mathrm{new}}_k$ to clients $k=1, \ldots, K$\;
	\For{each client $k \in \{1,\ldots,K\}$}{
		Generate privatized synthetic data for each variable as
		$\widetilde{X}_j^{\mathrm{priv}} = (\widetilde{F}_j^{\mathrm{priv}})^{-1} \big(\Phi(Z_j^{\mathrm{new}})\big)$.
	}
\end{algorithm}

\paragraph{The Iterated EVCDS (IEVCDS) Algorithm.}

The EVCDS method inevitably suffers from bias under MAR missingness because the dependency structure $\bm{\Omega}$ in the copula model is estimated from rank correlations, where the ranks are computed independently for each covariate using only observed values. To mitigate this bias, we impute missing entries via Gaussian latent variables: $\widetilde{X}_{j,\text{mis}} = (\widetilde{F}_j)^{-1} \big(\Phi(Z_{j,\text{mis}}^{'})\big)$. This yields a pseudo-complete dataset for each client with missingness, denoted by $\widetilde{\mathbf{X}}_k = (\mathbf{X}_{k,\text{obs}}, \widetilde{\mathbf{X}}_{k,\text{mis}})$. Based on this dataset, each client can compute complete local ranks, which are then privatized using the debiased randomized response (RR) mechanism and transmitted to the server.

The server aggregates these privatized complete ranks to estimate $\bm{\Omega}$. A latent sample is drawn from $\mathcal{N}(\mathbf{0}, \widetilde{\bm{\Omega}}^d)$ and subsequently adjusted according to $\widetilde{\mathbf{R}}^d$ to produce $\mathbf{Z}^{'}$. The partitioned matrices $\mathbf{Z}^{'}_k$ are then distributed back to the clients for estimating the marginal distributions ${F_j}$ with the aid of the perturbed complete ranks.

This process is iterated for $T$ rounds. If the privacy cost per iteration is $\epsilon_1$ for rank perturbation and $\epsilon_2$ for marginal privatization, the total privacy budget by the composition property of DP is $T\epsilon_1+\epsilon_2$ (Corollary~\ref{coro4}). Here, only the ranks shared with the server are perturbed at each iteration, while the margins used for imputation remain unprivatized, since the intermediate pseudo-complete datasets $\widetilde{\mathbf{X}}_k^{(t)}$ are never transmitted. The marginal privatization is performed only in the final iteration to ensure the synthetic data are privacy-preserving.

A technical detail concerns the initialization of $\mathbf{Z}^{'(0)}_{\text{mis}}$. We set $\widetilde{\mathbf{R}}^{d(0)} = \widetilde{\mathbf{R}}^d_{\text{obs}}$ and first obtain $\mathbf{Z}^{'(0)}_{\text{obs}}$ following Steps 7-14 in Algorithm~\ref{alg2}. The missing part $\mathbf{Z}^{'(0)}_{\text{mis}}$ is then sampled from the conditional normal distribution given $\mathbf{Z}^{'(0)}_{\text{obs}}$.
Specifically, for each subject $i$, sample $\mathbf{Z}_{i,\text{mis}}^{'} \sim \mathcal{N}(\bm{\mu}_{i,\text{mis}|\text{obs}}, \bm{\Omega}_{i,\text{mis}|\text{obs}})$, where 
$\bm{\mu}_{i,\text{mis}|\text{obs}} = \widetilde{\bm{\Omega}}^{d(0)}_{\mathcal{I}_{i,\text{mis}}, \mathcal{I}_{i,\text{obs}}}(\widetilde{\bm{\Omega}}^{d(0)}_{\mathcal{I}_{i,\text{mis}}, \mathcal{I}_{i,\text{obs}}})^{-1}\mathbf{Z}^{'}_{i,\text{obs}}$,
$\bm{\Omega}_{i,\text{mis}|\text{obs}} = \widetilde{\bm{\Omega}}^{d(0)}_{\mathcal{I}_{i,\text{mis}}, \mathcal{I}_{i,\text{mis}}} - \widetilde{\bm{\Omega}}^{d(0)}_{\mathcal{I}_{i,\text{mis}}, \mathcal{I}_{i,\text{obs}}}(\widetilde{\bm{\Omega}}^{d(0)}_{\mathcal{I}_{i,\text{obs}}, \mathcal{I}_{i,\text{obs}}})^{-1} \widetilde{\bm{\Omega}}^{d(0)}_{\mathcal{I}_{i,\text{obs}}, \mathcal{I}_{i,\text{mis}}}$,
with $\mathcal{I}_{i,\text{obs}}$ and $\mathcal{I}_{i,\text{mis}}$ denoting the observed and missing variable indices for subject $i$.

Finally, Theorem~\ref{thm4} analyzes the utility of synthetic data generated by Algorithm~\ref{alg3}. It shows that the discrepancy between the copula models fitted on the true dataset and on the privatized dataset converges to a stable limit determined jointly by the privatization mechanism and the missingness mechanism.

\begin{algorithm}[!h]
	\caption{The IEVCDS Algorithm}
	\scriptsize
	\label{alg3}
	\KwIn{Local raw data $\mathbf{X}_k$ for clients $k=1,2,\ldots,K$ and $Y$, and the initialization $\widetilde{\mathbf{X}}_{k,\text{mis}}^{(0)} = \emptyset.$}
	\KwOut{Privatized pseudo data $\widetilde{\mathbf{X}}_k^{\text{priv}}$ for clients $k=1,2,\ldots,K$ and $\widetilde{Y}^{\text{priv}}$}
	\For{each iteration $t\in \{1,\cdots,T\}$}{
		\For{each client $k \in \{1,\ldots,K\}$}{
			Form the pseudo-complete dataset $\widetilde{\mathbf{X}}_k^{(t)} = (\mathbf{X}_{k,\text{obs}}, \widetilde{\mathbf{X}}_{k,\text{mis}}^{(t-1)})$\;
			Perturb the ranks of each column of $\widetilde{\mathbf{X}}_k^{(t)}$ to obtain $\widetilde{\mathbf{R}}_k^{(t)}$, then apply the debiased RR mechanism to obtain $\widetilde{\mathbf{R}}_k^{d(t)}$\;
			Send $\widetilde{\mathbf{R}}_k^{d(t)}$ to the server\;
		}
		\textbf{Server side:} \\
		Aggregate rank matrices as $\widetilde{\mathbf{R}}^{d(t)} := (\widetilde{\mathbf{R}}_1^{d(t)}, \ldots, \widetilde{\mathbf{R}}_K^{d(t)})$\;
		Compute the copula correlation matrix estimator $\widetilde{\bm{\Omega}}^{d(t)}$ from $\widetilde{\mathbf{R}}^{d(t)}$\;
		Draw a sample $\mathbf{Z}^{(t)} \in \mathbb{R}^{N \times (p+1)}$ from $\mathcal{N}_{p+1}(\mathbf{0}, \widetilde{\bm{\Omega}}^{d(t)})$\;
		Sort $Z_0^{(t)}$ to match the order of $\widetilde{R}_0^{d(t)}$\;
		\For{$j = 1$ \KwTo $p$}{
			Sort $Z_j^{(t)}$ according to $\widetilde{R}_j^{d(t)}$\;
			\For{$i = 1$ \KwTo $N$}{
				Compute the truncated interval $(z_l^{(t)}, z_u^{(t)})$ as defined in Equation \eqref{Eq6}\;
				Compute conditional mean and variance:
				\begin{align*}
				\mu_{ij}^{(t)} &= \widetilde{\bm{\Omega}}^{d(t)}_{j,1:(j-1)} \big(\widetilde{\bm{\Omega}}^{d(t)}_{1:(j-1),1:(j-1)}\big)^{-1} \mathbf{z}^{(seq)},\\ 
				\sigma_j^{2(t)} &= \widetilde{\bm{\Omega}}^{d(t)}_{j,j} - \widetilde{\bm{\Omega}}^{d(t)}_{j,1:(j-1)} \big(\widetilde{\bm{\Omega}}^{d(t)}_{1:(j-1),1:(j-1)}\big)^{-1} \widetilde{\bm{\Omega}}^{d(t)}_{1:(j-1),j},
				\end{align*}
				where $\mathbf{z}^{(seq)} = \big(Z^{'(t)}_{i1}, \ldots, Z^{'(t)}_{i(j-1)}\big)$ and $Z^{'(t)}_{i0} = Z_{i0}^{(t)}$\;
				Sample $Z^{'(t)}_{ij} \sim \mathcal{N}\big(\mu_{ij}^{(t)}, \sigma_j^{2(t)}\big)\mathbb{I}\big(z_l^{(t)}, z_u^{(t)}\big)$ \;
			}
		}
		Obtain the adjusted latent sample matrix $\mathbf{Z}^{'(t)}$, then distribute the corresponding submatrices $\mathbf{Z}^{'(t)}_k$ back to each client $k$\;
		\For{each client $k \in \{1,\ldots,K\}$}{
			Estimate marginal distributions by
			$\widetilde{F}^{(t)}_j(\tilde{x}) = \Phi\big(\widetilde{Z}^{(t)}_j(\tilde{x})\big)$ and smooth it using monotone spline,
			where $\widetilde{Z}^{(t)}_j(\cdot)$ and $\tilde{x}$ are defined in Equations \eqref{Eq7} and \eqref{Eq8}\;
			Obtain pseudo values for missing variable as
			$\widetilde{X}_{j,\text{mis}}^{(t)} = (\widetilde{F}_j^{(t)})^{-1} \big(\Phi(Z_{j,\text{mis}}^{'(t)})\big)$.
		}	
	}
	\textbf{Server side:} \\
	Draw a new sample $\mathbf{Z}^{\mathrm{new}} \sim \mathcal{N}_{p+1}(\mathbf{0}, \widetilde{\bm{\Omega}}^{d(T)})$,  
	rearrange $\mathbf{Z}^{\mathrm{new}}$ following steps similar to lines 8--14, and distribute submatrices $\mathbf{Z}^{\mathrm{new}}_k$ to clients $k=1, \ldots, K$\;
	\For{each client $k \in \{1,\ldots,K\}$}{
		Approximate each $\widetilde{F}_j^{(T)}$ using privatized Bernstein polynomials:
		$\widetilde{F}_j^{(T)\mathrm{priv}}(\tilde{x}) = \widetilde{B}_l^{(h)}(\widehat{F}_j^{(T)})(\tilde{x})$\;
		Generate privatized synthetic data for each variable as
		$\widetilde{X}_j^{\mathrm{priv}} = (\widetilde{F}_j^{(T)\mathrm{priv}})^{-1} \big(\Phi(Z_j^{\mathrm{new}})\big)$.
	}
\end{algorithm}

\csection{Theoretical Justification}\label{sec4}
In this section, we establish the theoretical foundations of the privatized copula estimator constructed via the RR mechanism and the nonparametric approach. Building on these preliminaries, we then analyze the theoretical properties of Algorithms~1--3, focusing on both privacy and utility guarantees. The privacy guarantee represents the fundamental requirement of any differentially private algorithm, while the utility guarantee characterizes the effectiveness of the proposed methods relative to their non-private counterparts.  

\subsection{Preliminary Results}

\begin{theorem}\label{theorem1}
	Suppose there are $K$ clients in a vertical federated system, collectively holding $p+1$ variables and $N$ samples, with the variables uniformly distributed across the clients. Let $\epsilon_{1}$ denote the total privacy budget allocated to the rank perturbation step. 
	Define the noise parameter as $\theta = \left(1+\exp\left(\frac{\epsilon_{1}K}{p+1}\right)\right)^{-1}$. Then, with probability at least $1 - \delta$, the following bounds hold:
	\begin{align}
		\frac{\|\widehat{\bm{\Omega}} - \widetilde{\bm{\Omega}}\|_F}{\|\widehat{\bm{\Omega}}\|_F} & = \mathcal{O}_{\mathbb{P}}\left(\theta + \frac{\sqrt{p \log (1/\delta)}}{N}\right), \label{eq:first}\\
		\left\| \widehat{\bm{\Omega}} - \widetilde{\bm{\Omega}}^d \right\|_F &= \mathcal{O}_{\mathbb{P}} \left( \frac{(2\theta^2 - 2\theta + 1)p}{(1 - 2\theta)^2 N} \sqrt{ \log \left( \frac{2p^2}{1 - \delta} \right)} \right). \label{eq:second}
	\end{align}
\end{theorem}

Theorem \ref{theorem1} shows that $\widetilde{\bm{\Omega}}$ obtained from the original perturbed ranks is inconsistent, whereas $\widetilde{\bm{\Omega}}^{d}$ derived from the debiased perturbed ranks is consistent.
Let $a = \epsilon_{1}K/(p+1)$ and denote $f(a)=(e^{2a}+1)/(e^a-1)^2$, then the bound \eqref{eq:second} scales as $\mathcal{O}_{\mathbb{P}}\left((p/N)f(a)\sqrt{\log (2p^2/(1-\delta))}\right)$. As $a \rightarrow \infty$ (i.e. $\epsilon_{1}$ is large), $f(a) \rightarrow 1$ and the error reduces to the usual $\mathcal{O}_{\mathbb{P}}(p/N)$ rate (up to logarithmic factors). In contrast, when $a \rightarrow 0$ (i.e., $\epsilon_{1}$ is very small), $f(a)$ diverges like $2/a^2$, implying that the estimation error scales approximately as $\mathcal{O}_{\mathbb{P}}\left(p(p+1)^2/(N\epsilon_{1}^2K^2)\right)$ (up to $\sqrt{\log (\cdot)}$). 
Thus, in the strong-privacy regime, the bound has an $\epsilon_{1}^{-2}$ dependence, making it necessary to substantially increase $N$ to counteract the resulting amplification effect.

\begin{corollary}\label{corol-1}
	Assume that $\widehat{F}_j^{\mathrm{priv}}$ is the privatized version of $\widehat{F}_j$ obtained via the Iterated Bernstein operator in Eq.~\eqref{Eq9} with privacy parameter $\epsilon$. 
	If $\widehat{F}_j$ is $(2h, T)$-smooth, then with probability at least $1-\delta$, there exists 
	$l = \max \left\{1, \left(\frac{\epsilon N}{\log (1/\delta)}\right)^{\frac{1}{h+1}}\right\}$ such that
	\begin{equation*}
		\max_{x} \left|\widehat{F}_j^{\text{priv}}(x) - \widehat{F}_j(x)\right| = \mathcal{O}_{\mathbb{P}}\left(\frac{l+1}{N\epsilon}\log(1/\delta)\right)^{\frac{h}{h+1}}.
	\end{equation*}
\end{corollary}
Corollary \ref{corol-1} is a direct consequence of Theorem 3 in \cite{alda2017bernstein}, and its proof is omitted for brevity.

\begin{theorem}\label{theorem2}
	Suppose $\{Z_i\}_{i=1}^N \overset{\text{i.i.d.}}{\sim} \mathcal{N}(0,1)$ and define $\{X_i\}_{i=1}^N = \{G(Z_i)\}_{i=1}^N \sim F_X$, where $G$ is a monotone increasing function. For any $x \in \mathbb{R}$, define
	\begin{align*}
		\widetilde{Z}(x) = \max\left\{ \{Z_{i}: \widetilde{R}(X_{i})\leq \widetilde{R}(x)\} \cup \{Z_{i}: \widetilde{R}(X_{i}) = \min_{1 \le i \le N} \widetilde{R}(X_{i})\}, i=1,\cdots,N \right\}.
	\end{align*}
	where $R(\cdot)$ and $\widetilde{R}(\cdot)$ denote the true and perturbed rank functions, respectively. Define the privatized estimator of $F_X(x)$ as
	$\widetilde{F}(x) := \Phi(\widetilde{Z}(x))$. Denote $\theta = 1/(1 + e^{\epsilon})$. Then we have the following results:
	\begin{enumerate}
		\item For any $x \in \mathbb{R}$, it holds that
		\begin{equation}\label{Eq4}
			\left|\widetilde{F}(x) - F_X(x)\right| = \mathcal{O}_{\mathbb{P}}\left(\frac{1}{(1-2\theta)\sqrt{N}}\right).
		\end{equation}
		\item The function $\widetilde{F}(x)$ is not a valid cumulative distribution function in the strict sense. Let $\{\widetilde{X}_i\}_{i=1}^N$ denote the perturbed version of $\{X_i\}_{i=1}^N$, obtained by reordering $\{X_i\}_{i=1}^N$ according to their perturbed ranks. 
		Then, the function $\widetilde{F}(\widetilde{x})$ is a valid distribution function and satisfies: for any $x \in \mathbb{R}$,
		\begin{equation}\label{Eq5}
			\left|\widetilde{F}(\widetilde{x}) - F_X(x)\right| = \mathcal{O}_{\mathbb{P}}\left(\frac{1}{(1-2\theta)\sqrt{N}}\right),
		\end{equation}
		where $\widetilde{x} := X_{(\widetilde{R}(x))}$.
	\end{enumerate}
\end{theorem}

Theorem \ref{theorem2} states that the proposed privatized marginal estimator, constructed using the latent Gaussian variables and the perturbed ranks, is consistent, although its convergence rate is reduced by a privacy factor due to the added noise. Furthermore, Theorem \ref{theorem2-1} shows that the proposed marginal CDF estimator remains consistent under MAR missingness, maintaining the same convergence rate.

\begin{theorem}\label{theorem2-1}
	Let $\{\mathbf{Z}_i\}_{i=1}^N = \{(Z_{i1}, Z_{i2})\}_{i=1}^N \overset{i.i.d.}{\sim} \bm{\Phi}$, 
	where $\bm{\Phi}$ is a bivariate normal distribution with standard normal marginals $\Phi$. 
	Suppose $\{\mathbf{X}_i\}_{i=1}^N = \{(X_{i1}, X_{i2})\}_{i=1}^N = \left\{\big(F_1^{-1}(\Phi(Z_{i1})), \, F_2^{-1}(\Phi(Z_{i2}))\big)\right\}_{i=1}^N$,
	which has joint distribution function $F$ with marginals $F_1$ and $F_2$. Assume that $X_2$ is fully observed while $X_1$ is MAR. 	
	For $j \in \{1,2\}$, define
	{\small
		\begin{align*}
			\widetilde{Z}_j(x) 
			= \max \Bigg\{ \, \{Z_{ij}^{'\text{obs}} : \widetilde{R}(X_{ij}^{\text{obs}}) \leq \widetilde{R}(x)\} 
			\, \cup \, \{Z_{ij}^{'\text{obs}} : \widetilde{R}(X_{ij}^{\text{obs}}) = \min_{1 \leq i \leq N} \widetilde{R}(X_{ij}^{\text{obs}})\}, \, i=1,\dots,N \Bigg\}.
		\end{align*}
	}
	Denote the privatized marginal distribution of $X_1$ as $\widetilde{F}_1(x) := \Phi(\widetilde{Z}_1(x))$. Let $\theta = 1/(1+e^{\epsilon})$. Then, for all $x \in \mathbb{R}$, we have
	\begin{equation*}
		\left|\widetilde{F}_1(x) - F_1(x)\right|
		= \mathcal{O}_{\mathbb{P}}\!\left(\frac{1}{(1-2\theta)\sqrt{N}}\right).
	\end{equation*}
\end{theorem}

\begin{corollary}\label{corol-2}
	Define $\Delta \widetilde{F}_j := \sup_{X_j \sim X_j^{'}} \left|\widetilde{F}_j(\tilde{x}) - \widetilde{F}_j(\tilde{x}')\right|$, where $X_j$ and $X_j^{'}$ differ in at most one record. Let $\widetilde{F}_j^{\mathrm{priv}}$ denote the privatized version of $\widetilde{F}_j$ obtained via the Iterated Bernstein operator in Eq.~\eqref{Eq9} with privacy parameter $\epsilon$.
	Then $\Delta \widetilde{F}_j = O\left(\frac{1}{N}\right) $.
	If $\widetilde{F}_j$ is $(2h, T)$-smooth, then with probability at least $1-\delta$, there exists 
	$l = \max \left\{1, \left(\frac{\epsilon N}{\log (1/\delta)}\right)^{\frac{1}{h+1}}\right\}$ such that
	\begin{equation*}
		\max_{\tilde{x}} \left|\widetilde{F}_j^{\text{priv}}(\tilde{x}) - \widetilde{F}_j(\tilde{x})\right| = \mathcal{O}_{\mathbb{P}}\left(\frac{l+1}{N\epsilon}\log(1/\delta)\right)^{\frac{h}{h+1}}.
	\end{equation*}
\end{corollary}

\subsection{Privacy Guarantee and Utility Guarantee}

 First, we define Vertical Distributed Attribute Differential Privacy (VDADP) for vertical federated learning. The classical differential privacy \citep{dwork2006differential} is defined over pairs of datasets that differ in only one individual, i.e., with a Hamming distance of one. This formulation ensures that the output of a randomized algorithm does not significantly change with the inclusion or exclusion of any single individual, thereby protecting individual-level participation.
However, in the context of VFL, the set of individuals across all participating clients is aligned and fixed. Thus, it is not necessary to protect the presence or absence of individuals. Instead, the primary privacy concern lies in ensuring that the attributes of a particular individual held by one client cannot be inferred, even when all other attributes of that individual (and of all other individuals) held by other clients are known. Therefore, it is natural to extend the standard DP to VFL scenario as follows.

\begin{definition}[Vertical Distributed Attribute Differential Privacy]
	In a VFL system, define the attribute-level distance between two datasets $\mathbf{X}$ and $\mathbf{X}^{\prime}$ as:
	$$
		\Delta (\mathbf{X}, \mathbf{X}^{\prime}) = |\{(i, k): \mathbf{X}_{i,k} \neq \mathbf{X}_{i,k}^{\prime}, 1\leq i \leq N, 1 \leq k \leq K\}|.
	$$
	A randomized algorithm $\mathcal{A}: \mathcal{X}^N \rightarrow \mathbb{A}$ is $(\epsilon, \delta)$-vertical distributed attribute differentially private ($(\epsilon, \delta)$-VDADP) if for any pair of datasets $\mathbf{X} \in \mathcal{X}^N$ and $\mathbf{X}^{\prime} \in \mathcal{X}^N$ with $\Delta (\mathbf{X}, \mathbf{X}^{\prime}) = 1$, the following holds
	\begin{equation*}
		\mathbb{P}\{\mathcal{A}(\mathbf{X} \in A)\} \leq e^{\epsilon} \mathbb{P}\{\mathcal{A}(\mathbf{X}^{\prime} \in A)\} + \delta,
	\end{equation*}
for every measurable subset $A \in \mathbb{A}$.
\end{definition}
 The parameters $\epsilon$ and $\delta$ are privacy budgets indicating the strength of privacy protection from the mechanism. Smaller $\epsilon$ or $\delta$ indicates better privacy protection. When $\delta = 0$, the definition reduces to pure VDADP, which provides a stricter privacy guarantee.  Note that, when $K = 1$, $(\epsilon, \delta)$-VDADP recovers the classical $(\epsilon, \delta)$-DP. In this work, we focus on the pure privacy setting and adopt $\delta = 0$ throughout the analysis.

\begin{remark}
	Classical DP (i.e., sample-level DP) protects the presence or absence of an entire individual in the dataset. In contrast, VDADP in the VFL setting provides client-wise, attribute-level protection, ensuring that the local features of an individual held by a particular client cannot be inferred.	
	From the perspective of privacy strength, classical DP is stricter, as it safeguards the participation of the entire individual, whereas VDADP only protects a subset of features from a single client. Conceptually, VDADP can be viewed as a localized relaxation of classical DP: any mechanism satisfying classical DP automatically satisfies VDADP, but the converse does not hold.
	Moreover, through the composition theorem, multiple VDADP guarantees can be aggregated to achieve sample-level DP. Specifically, if each client applies an independent VDADP mechanism with privacy budget $\epsilon_k$, then the overall mechanism satisfies $(\sum_{k} \epsilon_k)$-DP at the sample level.
\end{remark}

\begin{theorem}[Differential Privacy of VCDS]\label{thm1}                                                                                                                                                                                                                                                                                                                                                                                                                                                                      
	The proposed VCDS mechanism in Algorithm \ref{alg1} returns a privacy-preserving dataset $\widetilde{\mathcal{D}}_k =  \widetilde{\mathbf{X}}_k^{\text{priv}}$ for each client $2\leq k \leq K$ and $\widetilde{\mathcal{D}}_{1} = (\widetilde{\mathbf{X}}_{1}^{\text{priv}}, \widetilde{Y}^{\text{priv}})$ for the server $k=1$. Suppose that for client $k$, the privacy budgets allocated to the rank perturbation and marginal distribution privatization steps are $\epsilon_{1k}$ and $\epsilon_{2k}$, respectively.
	Then the privatized dataset $\widetilde{\mathcal{D}}_k$ ($1\leq k \leq K$) satisfies $(\epsilon_{1k} + \epsilon_{2k})$-VDADP, and the aggregated privatized dataset across all clients, denoted by $\widetilde{\mathcal{D}} = \bigcup_{k=1}^{K} \widetilde{\mathcal{D}}_k$, satisfies $\sum_{k=1}^{K}(\epsilon_{1k} + \epsilon_{2k})$-DP.
\end{theorem}

Theorem~\ref{thm1} formally guarantees that the VCDS mechanism enforces rigorous privacy protection at both the local and global levels. Specifically, for each client $k$, a personalized privacy budget $\epsilon_k$ is partitioned into two components: $\epsilon_{1k}$ is dedicated to the perturbation of rank information, while $\epsilon_{2k}$ is allocated to the noise injection into the latent variables. This decomposition enables fine-grained control over privacy leakage in different stages of the data transformation process.

Under the VDADP framework, the privacy guarantee ensures that, even if an adversary possesses complete knowledge of all attributes of all individuals except for a single attribute vector locally stored by client $k$, the adversary cannot reliably infer the protected attribute from the privatized output. Additionally, the mechanism inherently protects against attribute inference and membership inference attacks, ensuring that the presence or absence of a specific value or sample has a limited impact on the sanitized output.

\begin{corollary}[Differential Privacy of EVCDS]\label{coro3}
The EVCDS mechanism described in Algorithm~\ref{alg2} satisfies $(\epsilon_{1k}+\epsilon_{2k})$-VDADP for each client $k$ ($1 \leq k \leq K$),  
where $\epsilon_{1k}$ and $\epsilon_{2k}$ denote the privacy budgets allocated to the rank perturbation step and the marginal distribution privatization step, respectively.  
For the entire VFL system, the EVCDS mechanism guarantees $\big(\sum_{k=1}^{K}(\epsilon_{1k}+\epsilon_{2k})\big)$-DP.
\end{corollary}

Corollary \ref{coro3} follows directly from the privacy guarantee of Algorithm~1 and the composition property of differential privacy. In Corollary \ref{coro4}, since the ranks of the completed variables are re-perturbed at each iteration, the privacy cost of rank perturbation accumulates. Consequently, the result is derived from the privacy guarantee of Algorithm~2 combined with the composition property of differential privacy.

\begin{corollary}[Differential Privacy of IEVCDS]\label{coro4}
	Suppose that for client $k$, the rank perturbation and marginal privatization incurs a privacy cost of $\epsilon_{1k}$ and $\epsilon_{2k}$, respectively. Then, at each iteration, Algorithm~\ref{alg3} satisfies $\epsilon_{1k}$-VDADP for client $k$ ($1 \leq k \leq K$).  
	After $T$ iterations, it outputs $\widetilde{\mathbf{X}}_k^{\text{priv}}$ for each client and $\widetilde{Y}^{\text{priv}}$ for the server, which together satisfy $\big(\sum_{k=1}^{K}(T\epsilon_{1k}+\epsilon_{2k})\big)$-DP.
\end{corollary}

\begin{theorem}[Utility of VCDS]\label{thm2} 
Let $(\mathbf{X}, Y)$ be a vertically partitioned dataset of $N$ i.i.d. samples, with $p$ covariates distributed across $K$ clients, where client $k$ contains $p_k$ covariates. 
Assume that $\mathbf{X}$ suffers from client-wise MCAR missingness with missing rate $\rho_k$ on client $k$, and denote $\rho_{\max} := \max_k \rho_k$.
Let $(\widetilde{\mathbf{X}}^{\mathrm{priv}}, \widetilde{Y}^{\mathrm{priv}})$ be the privatized version produced by the VCDS mechanism (Algorithm 1) under total privacy budget $\epsilon_1$ for rank perturbation and $\epsilon_2$ for marginal distribution privatization.

Let $d(\mathbf{x}, y)$ and $\tilde{d}(\mathbf{x}, y)$ denote the Gaussian copula densities fitted on the original full data and privatized data, respectively, with copula correlation matrices $\widehat{\bm{\Omega}}$ and $\widetilde{\bm{\Omega}}^d_{\text{obs}}$.

Assume $\widehat{\bm{\Omega}}$ is positive definite with $\lambda_{\min}(\widehat{\bm{\Omega}}) > \kappa > 0$. Let $h \in \mathbb{N}_{+}$. Denote $\theta = \left(1+\exp \left(\frac{K\epsilon_{1}}{p+1}\right)\right)^{-1}$, and $C(\theta) = \frac{2\theta^2-2\theta+1}{(1-2\theta)^2}$. denote 
$\rho_{\mathrm{I}} := \sum_{k=1}^{K}\frac{p_k^2}{1-\rho_k} + \sum_{k\neq l}\frac{p_kp_l}{(1-\rho_k)(1-\rho_l)}, \rho_{\mathrm{II}} := \sum_{k=1}^K \frac{p_k^2}{(1-\rho_k)^2} + \sum_{k\neq l}\frac{p_kp_l}{(1-\rho_k)^2(1-\rho_l)^2}$,
Then, with high probability at least $1-\delta$, the KL divergence between the two densities satisfies:
{\scriptsize 
	\begin{equation*}
		D_{KL}(d(\mathbf{x}, y) \,\|\, \tilde{d}(\mathbf{x}, y)) = \mathcal{O}_{\mathbb{P}}
		\left(\frac{\kappa^2 \rho_{\mathrm{I}}}{N} + \frac{\kappa^2C(\theta)^2\rho_{\mathrm{II}}}{N^2}\log \left(\frac{p}{\delta}\right) + \frac{p}{\sqrt{N}(1-\rho_{\max})} + \left(\frac{p^2}{NK(1-\rho_{\max})\epsilon_2}\log \left(\frac{1}{\delta}\right)\right)^{\frac{h}{h+1}}\right).
	\end{equation*}
}
\end{theorem}

Theorem~\ref{thm2} establishes a non-asymptotic upper bound on the Kullback-Leibler (KL) divergence between two Gaussian copula models: one fitted on the original full dataset, and the other on the privatized dataset generated by the proposed VCDS mechanism. 
This result quantifies the impact of privacy-preserving perturbations and MCAR missingness on the fidelity of the estimated data distribution.
The bound is mainly composed of four components: the estimation error of the rank correlation matrix due to missing data, the error introduced by rank perturbation, the marginal distribution estimation error due to finite sample size, and the error arising from marginal distribution privatization. As $N \rightarrow \infty$, the dominant terms vanish, indicating that VCDS remains effective in preserving data utility for large samples.
In particular, when $\rho_k = \rho$ and $p_k$ are identical across clients, the bound reduces to $\mathcal{O}_{\mathbb{P}}
\left(\frac{\kappa^2 p^2(K-\rho)}{KN(1-\rho)^2} + \frac{\kappa^2C(\theta)^2p^2[(1-\rho)^2+K-1] }{KN^2(1-\rho)^4}\log \left(\frac{p}{\delta}\right) + \frac{p}{\sqrt{N}(1-\rho)} + \left(\frac{p^2}{NK(1-\rho)\epsilon_2}\log \left(\frac{1}{\delta}\right)\right)^{\frac{h}{h+1}}\right)$.

This result confirms that the VCDS mechanism achieves a meaningful privacy-utility trade-off under fixed missing rates: the divergence between the estimated distributions decreases as (i) the privacy budgets $\epsilon_1$ and $\epsilon_2$ increase (i.e., less perturbation), or (ii) the sample size $N$ grows. In addition, the dependence on $\kappa$ shows that the bound tightens when the original copula correlation matrix is well-conditioned. Overall, the theorem demonstrates that VCDS provides a close approximation of the joint distribution under reasonable privacy budgets and sample sizes.

\begin{theorem}[Utility of EVCDS]\label{thm3}
	Consider a vertically partitioned dataset $(\mathbf{X}, Y)$ of $N$ i.i.d. samples with $p$ covariates distributed across $K$ clients, where client $k$ holds $p_k$ covariates. Suppose $\mathbf{X}$ is subject to client-wise MAR missingness, with missing rate $\rho_k$ on client $k$, and let $\rho_{\max} := \max_k \rho_k$. Denote by $\pi_{kl}\geq \varpi>0$ the probability that covariates on clients $k$ and $l$ $(k\neq l)$ are simultaneously observed.  
	
	Let $(\widetilde{\mathbf{X}}^{\mathrm{priv}}, \widetilde{Y}^{\mathrm{priv}})$ be the privatized data generated by the EVCDS mechanism (Algorithm~\ref{alg2}) under privacy budgets $\epsilon_1$ (rank perturbation) and $\epsilon_2$ (marginal privatization). Denote by $d(\mathbf{x},y)$ and $\tilde{d}(\mathbf{x},y)$ the Gaussian copula densities fitted on the original full data and privatized data, with correlation matrices $\widehat{\bm{\Omega}}$ and $\widetilde{\bm{\Omega}}^d$, respectively.  
	
	Assume $\widehat{\bm{\Omega}}$ is positive definite with $\lambda_{\min}(\widehat{\bm{\Omega}}) > \kappa > 0$. Let $h \in \mathbb{N}_{+}$. Denote $\theta = \left(1+\exp \left(\frac{K\epsilon_{1}}{p+1}\right)\right)^{-1}$, and $C(\theta) = \frac{2\theta^2-2\theta+1}{(1-2\theta)^2}$. 
	Denote $\rho_{\mathrm{I}}^{'} := \sum_{k=1}^{K}\frac{p_k^2}{1-\rho_k} + \sum_{k\neq l}\frac{p_kp_l}{\pi_{kl}}, \rho_{\mathrm{II}}^{'} := \sum_{k=1}^K \frac{p_k^2}{(1-\rho_k)^2} + \sum_{k\neq l}\frac{p_kp_l}{\pi_{kl}^2}$.
	Define $\mathcal{E}_{\mathrm{marg}} := \left( \frac{p \log(1/\delta)}{NK(1-\rho_{\max})\epsilon_2} \right)^{\frac{h}{h+1}} + \frac{1}{(1-2\theta)\sqrt{N(1-\rho_{\max})}}$.
	Then, with high probability at least $1-\delta$, the KL divergence between the two densities satisfies:
	\begin{equation*}
		D_{KL}(d(\mathbf{x}, y) \,\|\, \tilde{d}(\mathbf{x}, y)) \leq \mathcal{O}_{\mathbb{P}} \left( \frac{\kappa^2\rho_{\mathrm{I}}^{'}\log p}{N} + \kappa^2 B^2_{MAR} + \frac{\kappa^2 C(\theta)^2\rho_{\mathrm{II}}^{'}}{N^2}\log \left(\frac{p}{\delta}\right) + p\mathcal{E}_{\mathrm{marg}} \right).
	\end{equation*}
\end{theorem}

Theorem~\ref{thm3} presents a utility guarantee for EVCDS under client-wise MAR missingness and differential privacy constraints. Specifically, it shows that the KL divergence between the copula density estimated from the original full data (without missingness) and the privatized data is bounded by four terms:
(1) sampling error term of order $\frac{\rho_{\mathrm{I}}'\log p}{N}$;
(2) a bias term $B_{MAR}^2$ caused by the MAR mechanism when estimating the correlation based on the observed data, its explicit formula is given in Appendix;
(3) a higher-order variance term $\frac{\rho_{\mathrm{II}}'C(\theta)^2}{N^2}$ due to privacy-induced perturbations, where $C(\theta)$ is the variance inflation factor associated with debiased randomized response;
(4) a marginal error term $\mathcal{E}_{\mathrm{marg}}$ arising from privatization of the marginal distributions.
While the first, third, and fourth terms decay with larger sample size or looser privacy budgets, the MAR bias term persists and represents an intrinsic limitation of learning under MAR. Thus, EVCDS guarantees asymptotic fidelity up to the MAR bias, which cannot be eliminated without additional measures.

\begin{theorem}[Utility of IEVCDS]\label{thm4}
	Consider a vertically partitioned dataset $(\mathbf{X}, Y)$ of $N$ i.i.d. samples with $p$ covariates distributed across $K$ clients, where client $k$ holds $p_k$ covariates. Suppose $\mathbf{X}$ is subject to client-wise MAR missingness, with missing rate $\rho_k$ on client $k$, and let $\underline{\pi} := 1- \max_k \rho_k$. Denote by $\pi_{kl}\geq \varpi>0$ the probability that covariates on clients $k$ and $l$ $(k\neq l)$ are simultaneously observed.  
	
	Let $(\widetilde{\mathbf{X}}^{\mathrm{priv}}, \widetilde{Y}^{\mathrm{priv}})$ be the privatized data generated by the IEVCDS mechanism (Algorithm \ref{alg3}) under privacy budgets $\epsilon_1$ (rank perturbation) and $\epsilon_2$ (marginal privatization). Denote by $d(\mathbf{x},y)$ and $\tilde{d}(\mathbf{x},y)$ the Gaussian copula densities fitted on the original full data and privatized data, with correlation matrices $\widehat{\bm{\Omega}}$ and $\widetilde{\bm{\Omega}}^{d(T)}$, respectively.  
	
	Assume $\widehat{\bm{\Omega}}$ is positive definite with $\lambda_{\min}(\widehat{\bm{\Omega}}) > \kappa > 0$. Let $h \in \mathbb{N}_{+}$. 
	Suppose Algorithm 3 runs for $T$ iterations. 
	Denote $\theta = \left(1+\exp \left(\frac{K\epsilon_{1}}{T(p+1)}\right)\right)^{-1}$, and $C(\theta) = \frac{2\theta^2-2\theta+1}{(1-2\theta)^2}$. 
	Then, with high probability at least $1-\delta$, the KL divergence between the two densities is upper bounded as:
	{\small
		\begin{equation*}
			D_{KL}(d(\mathbf{x}, y) \,\|\, \tilde{d}(\mathbf{x}, y)) \leq \mathcal{O}_{\mathbb{P}}\left(\frac{\kappa^2C(\theta)^2p^2}{(1-\alpha_{\bm{\Omega}})^2N^2\varpi^2}\log\left(\frac{p}{\delta}\right) + \frac{p}{1-\alpha_F} 
			\left(\frac{\sqrt{\frac{\log (1/\delta)}{N \underline{\pi}}} }{(1-2\theta)}
			+ \left(\frac{p \log \big(\frac{1}{\delta}\big)}{NK\epsilon_2}\right)^{\tfrac{h}{h+1}}\right)\right),
		\end{equation*}
	}
	where $\alpha_{\bm{\Omega}}$ and $\alpha_F$ are contraction coefficients satisfying $0<\alpha_{\bm{\Omega}}<1$ and $0<\alpha_F<1$.
\end{theorem}

Compared with EVCDS (Theorem~\ref{thm3}), the key difference lies in how IEVCDS addresses the MAR-induced bias. In EVCDS, the correlation matrix is estimated once from the observed ranks, which introduces a persistent MAR bias term $B_{MAR}^2$ that cannot be eliminated. IEVCDS mitigates this limitation by iteratively imputing missing values using latent Gaussian variables and updating the perturbed ranks in each iteration, effectively generating pseudo-complete datasets that allow more accurate correlation estimation.
The KL divergence bound in Theorem~\ref{thm4} reflects this improvement. Unlike EVCDS, where the MAR bias contributes a non-vanishing term, the IEVCDS bound depends on contraction coefficients $\alpha_{\bm{\Omega}}$ and $\alpha_F$ associated with the iterative adjustment procedure. These coefficients capture the shrinkage effect of repeated imputation and rank updates, ensuring that the discrepancy between the privatized and full-data copula densities decreases with the number of iterations $T$. Moreover, the other terms, corresponding to privacy-induced perturbations and marginal errors, are similar to EVCDS but scaled by $1/T$ factors in the rank perturbation, reflecting that repeated iterations distribute the privacy budget across multiple updates.

\section{Application to GLM Parameter Estimation with Variable Selection}\label{sec1}

In this section, we consider the problem of parameter estimation and variable selection for the VFL GLM model under privacy constraints and client-wise missingness. This is a challenge that has not been addressed in the literature to our knowledge. In the classical VFL setting, although raw data are not directly shared when optimizing the GLM model, the exchange of linear embeddings of the original data (e.g., $\mathbf{x}_i^{k\top}\bm{\beta}_k$) exposes participants to potential risks of privacy leakage.
A mainstream approach to addressing this problem is the application of differential privacy (DP) techniques. By introducing random noise into local parameters \citep{hu2020learning} or into local embeddings \citep{wang2020hybrid}, the associated privacy risks can be possibly mitigated. 
However, these methods are not designed to address the client-wise missingness problem. In addition, applying noise post hoc to the outputs of non-private variable selection procedures fails to preserve the statistical properties necessary for selection consistency, highlighting the need for specialized privacy-aware regularization mechanisms. In the following, we investigate the utility of the proposed privatized data for parameter estimation and variable selection in the VFL GLM model, beginning with the development of the estimation procedure.

Assume that the response variable $Y$ follows a generalized linear model (GLM):  
\begin{equation}\label{eq1}
	\mathbb{E}(Y|\mathbf{X}) = g(\bm{\beta}^{\top}\mathbf{X}) = g\left(\sum^K_{k=1}\mathbf{X}_k \bm{\beta}_k\right),
\end{equation}
where $\bm{\beta} = (\beta_1, \beta_2, \dots, \beta_p)^{\top}$ is the vector of unknown regression coefficients, and $\bm{\beta}_k \in \mathbb{R}^{p_k}$ denotes the subset of coefficients associated with the covariates held by client $k$. The function $g(\cdot)$ represents a known link function; for instance, $g(a) = a$ corresponds to the Gaussian linear model, whereas $g(a) = e^a/(1+e^a)$ characterizes the logistic regression model.

Here we consider the problem of parameter estimation and variable selection of GLM under VFL framework, assuming the complete and privatized data. Suppose that $y_i|\mathbf{x}_i$ has a density in the exponential class with the form
\begin{equation*}\label{eq2}
	p(y_i|\mathbf{x}_i) = \tilde{c}\exp \left\{\frac{y_i\mathbf{x}_i^{\top}\bm{\beta}^* - \psi(\mathbf{x}_i^{\top}\bm{\beta}^*)}{a(\tau)}\right\} = \tilde{c}\exp \left\{\frac{y_i\sum_{k=1}^{K}\mathbf{x}_i^{k\top}\bm{\beta}_k^* - \psi(\sum_{k=1}^{K}\mathbf{x}_i^{k\top}\bm{\beta}_k^*)}{a(\tau)}\right\},
\end{equation*}
where $\tilde{c}$ and $a(\tau)$ are scale constants, $\bm{\beta}^* = (\bm{\beta}_1^{*\top}, \cdots, \bm{\beta}_K^{*\top})$ denotes the true sparse model parameter vector. $\psi(\cdot)$ is the integral of the link function $g(\cdot)$ defined in \eqref{eq1}. We define $S = \{1 \leq j \leq p: \beta_j^* \neq 0\}$
as the support of $\bm{\beta}^*$ and $s = |S|$ as the sparsity level. To estimate $\bm{\beta}^*$, the most straightforward estimator is the solution of the (empirical) negative log-likelihood function
\begin{equation}\label{eq3}
	f(\bm{X}, \bm{y}, \bm{\beta}) = -\frac{1}{N}\sum_{i=1}^{N}y_i \sum_{k=1}^{K}\mathbf{x}_i^{k\top}\bm{\beta}_k + \frac{1}{N}\sum_{i=1}^{N}\psi\left(\sum_{k=1}^{K}\mathbf{x}_i^{k\top}\bm{\beta}_k\right).
\end{equation}
Furthermore, to pursue the sparsity structure of $\bm{\beta}^*$, we adopt a regularization term on the parameters and obtain the sparse estimated parameters as
\begin{equation*}
	\widehat{\bm{\beta}} = \arg \min_{\bm{\beta} \in \mathbb{R}^p} f(\bm{X}, \bm{y}, \bm{\beta}) + P_{\lambda}(|\bm{\beta}|) ,
\end{equation*}
where $\lambda > 0$ is the regularization hyper-parameter, and $P_{\lambda}(|\bm{\beta}|)=\sum_j P_{\lambda}(|\beta_j|)$ is a general folded concave penalty defined on $\beta \in (-\infty,\infty)$ satisfying: (i) $P_{\lambda}(\beta)$ is increasing and concave in $\beta \in[0,\infty)$ with $P_{\lambda}(0) = 0$; (ii) $P_{\lambda}(\beta)$ is differentiable in $\beta \in (0,\infty)$ with $P^{'}_{\lambda}(0):= P^{'}_{\lambda}(0+)\geq a_1 \lambda$; (iii) $P^{'}_{\lambda}(\beta) \geq a_1 \lambda$ for $\beta \in (0,a_2\lambda]$; (iv) $P^{'}_{\lambda}(\beta) = 0$ for $\beta \in [a\lambda, \infty)$ with the pre-specified constant $a>a_2$, where $a_1$ and $a_2$ are two fixed positive constants. The definition includes the two widely used nonconvex penalties: the smoothly clipped absolute deviation (SCAD; \citealp{fan2001variable}) and the minimax concave penalty (MCP; \citealp{zhang2010nearly}). 
For SCAD, its first derivative is defined as
\begin{equation*}
	P^{\prime}_{\lambda}(|\beta|) = \lambda \{I(|\beta|\leq \lambda) + \frac{(a\lambda - |\beta|)_{+}}{(a-1)\lambda}I(|\beta|>\lambda)\}
\end{equation*}
with $P^{\prime}_{\lambda}(0):= P^{\prime}_{\lambda}(0+) = \lambda$ and $a = 3.7$ as suggested in \citep{fan2001variable}. For MCP, 
its first derivative takes the form:
\begin{equation*}
	P^{\prime}_{\lambda}(|\beta|) = \frac{1}{a}(a\lambda - |\beta|)_{+}, \quad a > 1.
\end{equation*}
It is easy to see that $a_1=a_2=1$ for the SCAD, and $a_1 = 1-a^{-1}, a_2=1$ for the MCP.
To facilitate optimization, we adopt the local linear approximation (LLA; \citealp{zou2008one}) strategy, which
approximate the penalty function $P_{\lambda}(|\beta_j|)$ by first order Taylor's series expansion centered at
$\bm{\beta}^{(t)}$, the updated estimate from the $t$-th step in the course of iterations:
\begin{equation*}
	P_{\lambda}(|\beta_j|) \approx P_{\lambda}(|\beta_j^{(t)}|) + P_{\lambda}^{\prime}(|\beta_j^{(t)}|)(|\beta_j| - |\beta_j^{(t)}|).
\end{equation*}
Using this approximation, the nonconvex objective function resembles a reweighted $\ell_1$ penalized objective function. 

In the VFL context, the covariates are distributed across $K$ clients such that each client $k$ holds only $\mathbf{X}_k$. We introduce an auxiliary variable $\bm{\eta}$ to coordinate the computation across clients, leading to the following constrained optimization problem:
\begin{equation}\label{eq5}
	\mathcal{L}(\bm{\beta}) = -\frac{1}{N}\mathbf{y}^{\top}\sum_{k=1}^{K}\mathbf{X}_k\bm{\beta}_k + \frac{1}{N}\sum_{i=1}^{N} \psi\left(\sum_{k=1}^{K}\mathbf{x}_i^{k\top}\bm{\beta}_k\right) + \lambda \sum_{k=1}^{K}\|\bm{\alpha}_k \circ \bm{\beta}_k \|_1,~ s.t.~ \sum_{k=1}^{K}\mathbf{X}_k\bm{\beta}_k = \bm{\eta},
\end{equation}
where $\bm{\alpha}_k$ is an adaptive weight vector with elements $\alpha_{kj} = \lambda^{-1}P^{\prime}_{\lambda}(|\beta_{kj}^{(t)}|) \geq 0$, and $\circ$ denotes the Hadamard (element-wise) product. The auxiliary variable $\bm{\eta}$ corresponds to the aggregated intermediary results $\sum_{k=1}^{K}\mathbf{X}_k\bm{\beta}_k$ derived from the $K$ clients.


Typically, the above problem \eqref{eq5} can be solved by adopting ADMM based algorithm on its augmented Lagrangian:
{\small
\begin{equation*}
	\mathcal{L}_{\phi}(\bm{\beta},\bm{\eta};\bm{\gamma}) = -\frac{1}{N}\bm{y}^{\top}\bm{\eta} + \frac{1}{N}\sum_{i=1}^{N}\psi(\eta_i) + \lambda \sum_{k=1}^{K}\|\bm{\alpha}_k \circ \bm{\beta}_k \|_1 + \bm{\gamma}^{\top}\left( \sum_{k=1}^{K}\mathbf{X}_k\bm{\beta}_k - \bm{\eta}\right) + \frac{\phi}{2}\left\|\sum_{k=1}^{K}\mathbf{X}_k\bm{\beta}_k - \bm{\eta} \right\|^2_2,
\end{equation*}
}
where $\bm{\gamma}$ is the Lagrange multiplier and $\phi$ is the penalty parameter. 
The augmented Lagrangian function $\mathcal{L}_{\phi}(\bm{\beta},\bm{\eta};\bm{\gamma})$ can be recursively minimized to obtain $\bm{\beta}$, $\bm{\eta}$, and $\bm{\gamma}$ in a distributed manner.
Specifically, at $(t+1)$-th iteration, $\bm{\beta}$, $\bm{\eta}$, and $\bm{\gamma}$ can be sequentially updated as
{\small
\begin{align}
	\widehat{\bm{\beta}}_k^{(t+1)} &= \arg \min _{\bm{\beta}_k} \mathcal{L}_{\phi}\left(\bm{\beta}_k, \{\widehat{\bm{\beta}}_{k'}^{(t)}\}_{k'\neq k}, \widehat{\bm{\eta}}^{(t)}; \widehat{\bm{\gamma}}^{(t)}\right) \nonumber \\
	& = \arg \min _{\bm{\beta}_k}~ \lambda \|\bm{\alpha}_k \circ \bm{\beta}_k\|_1 + \langle \widehat{\bm{\gamma}}^{(t)}, \mathbf{X}_k\bm{\beta}_k\rangle + \frac{\phi}{2}\left\|\sum_{k'=1, k'\neq k}^{K}\mathbf{X}_{k'}\widehat{\bm{\beta}}_{k'}^{(t)} + \mathbf{X}_k\bm{\beta}_k - \widehat{\bm{\eta}}^{(t)}\right\|_2^2, \label{eq6-1} \\
	\bm{\eta}^{(t+1)} &= \arg \min _{\bm{\eta}} \mathcal{L}_{\phi}\left(\widehat{\bm{\beta}}^{(t+1)},\bm{\eta}; \widehat{\bm{\gamma}}^{(t)}\right) \nonumber \\
	&= \arg \min _{\bm{\eta}} ~ -\frac{1}{N}\bm{y}^{\top}\bm{\eta} + \frac{1}{N}\sum_{i=1}^{N}\psi(\eta_i) - \widehat{\bm{\gamma}}^{(t)\top}\bm{\eta} + \frac{\phi}{2}\left\|\sum_{k=1}^{K}\mathbf{X}_k \widehat{\bm{\beta}}_k^{(t+1)} - \bm{\eta}\right\|_2^2, \label{eq6-2}\\
	\widehat{\bm{\gamma}}^{(t+1)} &= \widehat{\bm{\gamma}}^{(t)} + \phi \left(\sum_{k=1}^{K}\mathbf{X}_k \widehat{\bm{\beta}}_k^{(t+1)} - \widehat{\bm{\eta}}^{(t+1)}\right). \label{eq6-3} 
\end{align}
}
The above updating formulas imply that all the $K$ clients can compute $\bm{\beta}_1,\cdots,\bm{\beta}_K$ locally in parallel. Then the server updates auxiliary variable $\bm{\eta}$ and dual variable $\bm{\gamma}$ by collecting intermediary results $\bm{\zeta}_k = \mathbf{X}_k\bm{\beta}_k$.

Algorithm \ref{alg4} elaborates the implementation for the ADMM-based federated minimization of \eqref{eq5}, referred to as VFL-GLM-VS (Vertical Federated Learning for Generalized Linear Models with Variable Selection). The model parameters can be computed locally for each client based on the received auxiliary variable $\bm{\eta}$ and dual variable $\bm{\gamma}$ using Eq. \eqref{eq6-1}. When $\bm{\beta}_k$ is obtained, the client
only needs to exchange the aggregated results $\bm{\zeta}_k = \mathbf{X}_k\bm{\beta}_k$ with the server. 
With the aggregated results $\{\bm{\zeta}_k\}^K_{k=1}$ collected from the $K$ clients, the server is able to update $\bm{\eta}$ and $\bm{\gamma}$, which are sent back to clients for subsequent update at the next iteration. 
This iterative process continues until both the primal residual $\mathbf{r}^{t} = \sum_{k=1}^{K}\mathbf{X}_k \bm{\beta}_k^{(t)} - \bm{\eta}^{(t)}$ and the dual residual $\mathbf{s}^{(t)} = \phi (\bm{\eta}^{(t)} - \bm{\eta}^{(t-1)})$ fall below their respective tolerances, i.e., $\|\mathbf{r}^{(t)}\|_2 \leq \varepsilon^{\text{pri}}$, $\|\mathbf{s}^{(t)}\|_2 \leq \varepsilon^{\text{dual}}$. Specifically, $\varepsilon^{\text{pri}} = \sqrt{N}\varepsilon^{\text{abs}} + \varepsilon^{\text{rel}}\max\{\|\sum_{k=1}^{K}\mathbf{X}_k\bm{\beta}_k^{(t)}\|_2, \|\bm{\eta}^{(t)}\|_2\}$, and $\varepsilon^{\text{dual}} = \sqrt{p}\varepsilon^{\text{abs}} + \varepsilon^{\text{rel}}\|\bm{\gamma}^{(t)}\|_2$. In practice, the absolute and relative tolerances can be set as  $\varepsilon^{\text{abs}}=\varepsilon^{\text{rel}}=0.001$.

\begin{algorithm}[!h]
	\caption{The VFL-GLM-VS Algorithm}
	\label{alg4}
	\KwIn{$\widehat{\bm{\beta}}^{(0)}=\bm{0}, \widehat{\bm{\eta}}^{(0)}=\bm{0}, \widehat{\bm{\gamma}}^{(0)}=\bm{0}$}
	\KwOut{$\widehat{\bm{\beta}}$}
	\While{the stopping criterion is not satisfied}{
		\For{client $k=1,\cdots,K$}{
			Computes $\sum_{k^{\prime} \neq k}\mathbf{X}_{k^{\prime}}\widehat{\bm{\beta}}_{k^{\prime}}^{(t)} - \widehat{\bm{\eta}}^{(t)} = \mathbf{h}^{(t)} - \mathbf{X}_{k}\widehat{\bm{\beta}}_{k}^{(t)}$\;
			Obtain $\widehat{\bm{\beta}}_k^{(t+1)}$ based on Eq. \eqref{eq6-1}\;
			Sends $\bm{\zeta}_k^{(t+1)} = \mathbf{X}_k \widehat{\bm{\beta}}_k^{(t+1)}$ to the server\;
		}
		Server computes $\widehat{\bm{\eta}}^{(t+1)}$ according to \eqref{eq6-2}\;
		Server computes $\widehat{\bm{\gamma}}^{(t+1)}$ according to \eqref{eq6-3}\;
		Server broadcasts $\mathbf{h}^{(t+1)}=\sum_{k=1}^{K}\bm{\zeta}^{(t+1)}_k - \widehat{\bm{\eta}}^{(t+1)}$ and $\widehat{\bm{\gamma}}^{(t+1)}$ to all clients.
	}
\end{algorithm}


\begin{theorem}\label{vs-thm1}
	Consider the GLM model with true $s$-sparse parameter $\bm{\beta}^*$ under vertical federated context. Let $\widehat{\bm{\beta}}$ be the estimator obtained by VFL-GLM-VS algorithm \ref{alg4} that satisfies KKT conditions. 
	Define $\widehat{S} = \{1\leq j\leq p: \hat{\beta}_j \neq 0\}$.
	Under Assumptions 1 in the Appendix, if $\lambda \ge 2\sigma\sqrt{\tfrac{2\log p}{N}}$ and the signal satisfies $\beta_{\min} := \min_{j\in S}|\beta_j^*| \ge \; \frac{4 C_1 \sqrt{s}\,\lambda}{\kappa_{\min}}$ for a suitable constant $C_{\mathrm{sig}}$ where $C_1\ge 1$ is any fixed constant.
	Then with probability at least $1-p^{-c_0}$, we have
	\begin{enumerate}
		\item (Estimation error) $\|\widehat{\bm{\beta}}-\bm{\beta}^*\|_2 \le \frac{2C_1\sqrt{s}\,\lambda}{\kappa_{\min}}$.
		\item (Support recovery) $\Pr(\widehat S=S)\to 1$. 
		\item (Asymptotic normality) If additionally $\max_{j\in S}\alpha_j \le \varepsilon_n$ with $\varepsilon_n \to 0$ and $\sqrt{N}\varepsilon_n \to 0$, then
		$\sqrt{N}(\widehat{\bm{\beta}}_S-\bm{\beta}_S^*) \overset{d}{\longrightarrow} \mathcal{N}(0,\mathbf{I}_S^{-1})$.
	\end{enumerate}
\end{theorem}

\begin{theorem}\label{vs-thm2}
	Let $(\mathbf{X},Y)$ be the original data satisfying $Y\mid X\sim \text{GLM}(\bm{\beta}^*)$ and let $(\widetilde{\mathbf{X}}^{\mathrm{priv}},\widetilde{Y}^{\mathrm{priv}})$ be the privatized data produced by the VCDS/IEVCDS mechanism. Suppose the fitted joint copula densities satisfy $D_{KL}\big(d(\mathbf{X},Y) \,\|\, \tilde d(\widetilde{\mathbf{X}}^{\mathrm{priv}},\widetilde{Y}^{\mathrm{priv}})\big) \le \varepsilon_N$,
	with $\varepsilon_N\to0$. Assume Assumption 1 in the Appendix holds for the original model and that the same RSC/Hessian bounds hold locally for the privatized population.
	
	Let $\widehat{\bm{\beta}}^{\mathrm{priv}}$ be any output of VFL-GLM-VS on the privatized data that satisfies the KKT conditions. 
	Set $\widehat S^{\mathrm{priv}}=\{j:\hat\beta_j^{\mathrm{priv}}\neq0\}$. 
	Take $\lambda \ge 2\sigma\sqrt{\tfrac{2\log p}{N}} + C_{\mathrm{priv}}\sqrt{\varepsilon_N}$,
	where $\sigma$ is from (A0) and $C_{\mathrm{priv}}$ is a constant depending on the Lipschitz properties of the score. If the signal obeys
	$\beta_{\min}:=\min_{j\in S}|\beta_j^*| \ge \frac{4C_1\sqrt{s}\,\lambda}{\kappa_{\min}}$
	for some constant $C_1\ge 1$, 
	then with probability at least $1-p^{-c_0}$ the estimator $\widehat{\bm{\beta}}^{\mathrm{priv}}$ satisfies
	\begin{enumerate}
		\item (Estimation error) $\|\widehat{\bm{\beta}}^{\mathrm{priv}}-\bm{\beta}^*\|_2 \le \dfrac{2C_1\sqrt{s}\,\lambda}{\kappa_{\min}}$.
		\item (Support recovery) $\Pr(\widehat S^{\mathrm{priv}}=S)\to 1$ as $N\to\infty$.
		\item (Asymptotic normality) If furthermore $\max_{j\in S}\alpha_j = o(N^{-1/2})$, then
		$\sqrt{N}(\widehat{\bm{\beta}}^{\mathrm{priv}}_S-\bm{\beta}_S^*) \overset{d}{\longrightarrow} \mathcal{N}(0,(\mathbf{I}_S^{\mathrm{priv}})^{-1})$,
		where $\mathbf{I}_S^{\mathrm{priv}}$ is the Fisher information matrix under the privatized population; in particular $\mathbf{I}_S^{\mathrm{priv}}\to \mathbf{I}_S$ as $\varepsilon_N\to0$.
	\end{enumerate}
\end{theorem}

Following \cite{hu2020learning}, the VFL-GLM-VS algorithm \ref{alg4} converges to a primal-dual stationary point. Let $\widehat{\bm{\beta}}$ denote the estimator obtained on the full, nonprivate dataset, and $\widehat{\bm{\beta}}^{\mathrm{priv}}$ the estimator obtained on the privatized dataset via the VCDS/IECVDS mechanisms. Theorem \ref{vs-thm1} guarantees that $\widehat{\bm{\beta}}$ achieves estimation consistency, variable selection consistency, and asymptotic normality, in accordance with classical high-dimensional theory. Theorem \ref{vs-thm2} shows that, if the KL divergence between the copula densities of the original and privatized data is bounded by $\varepsilon_N$, then by choosing the regularization parameter $\lambda$ to account for sample size $N$, dimension $p$, and $\varepsilon_N$, the privatized estimator $\widehat{\bm{\beta}}^{\mathrm{priv}}$ retains the same statistical guarantees. In particular, substituting the bounds on $\varepsilon_N$ derived in Theorems \ref{thm2} and \ref{thm4} yields explicit rates under both MCAR and MAR missingness.

\csection{SIMULATION}\label{sec3}

In this section, we assess the statistical properties of the proposed data privatization methods through comprehensive numerical experiments. We first generate privatized datasets from simulated original data, and then compare the results of parameter estimation and variable selection obtained from the privatized datasets with those from the original data. It is worth noting that the proposed algorithms, VCDS and EVCDS, can also be applied to complete datasets without missing values. Since there are currently no existing methods capable of addressing the VFL missingness problem under privacy constraints, our simulation study is organized into two parts. In the first part \ref{sec5-1}, we evaluate the performance of VCDS and EVCDS against competing methods on complete datasets; in the second part \ref{sec5-2}, we investigate the performance of the proposed methods under MCAR and MAR missingness mechanisms.

\subsection{Simulation Studies on Fully Observed Data}\label{sec5-1}

\paragraph{Data Generation} We generate $N$ samples $\{(\mathbf{x}_i, y_i)\}_{i=1}^N$, where the $p$-dimensional covariates $\mathbf{x}_i$ are distributed across $K$ clients. To simulate mixed-type data with complex dependencies, we first sample latent continuous variables $\mathbf{U} \in \mathbb{R}^{N \times p}$ from a finite mixture of multivariate normal distributions, and then transform $\mathbf{U}$ into different data types. 

Specifically, let $\mathbf{u}_i \stackrel{i.i.d.}{\sim} \sum_{j=1}^{3} \pi_j \mathcal{N}_p(\bm{\mu}_j, \bm{\Sigma}_j)$, where the mixing weights are $\bm{\pi} = (0.4, 0.3, 0.3)^\top$, and the component means are $\bm{\mu}_1 = \mathbf{0}_p$, $\bm{\mu}_2 = -\mathbf{1}_p$, and $\bm{\mu}_3 = \mathbf{1}_p$. The component covariance matrices are defined as follows:
(i) $\bm{\Sigma}_1$: off-diagonal elements are 0.1 for $k \neq k' \in \{1,\dots,K\}$; diagonals and off-diagonals within blocks are 1 and 0.3, respectively.
(ii) $\bm{\Sigma}_2$: a Toeplitz matrix with entries $\sigma_{j_1j_2} = 0.5^{|j_1 - j_2|}$.
(iii) $\bm{\Sigma}_3$: an AR(2)-type structure with $\sigma_{j_1j_2} = 0.5$ if $|j_1-j_2| = 1$, $0.25$ if $|j_1-j_2| = 2$, $1$ if $j_1 = j_2$, and $0$ otherwise.

Given the latent matrix $\mathbf{U}$, we construct a mixed-type covariate matrix $\mathbf{X} = (\mathbf{X}_1, \mathbf{X}_2, \mathbf{X}_3, \mathbf{X}_4) \in \mathbb{R}^{N \times p}$, where $\mathbf{X}_1 \in \mathbb{R}^{N \times q_1}$, $\mathbf{X}_2 \in \mathbb{R}^{N \times q_2}$, $\mathbf{X}_3 \in \mathbb{R}^{N \times q_3}$, and $\mathbf{X}_4 \in \mathbb{R}^{N \times q_4}$ correspond to continuous, multinomial, ordinal, and binary covariates, respectively. 
The generation mechanisms for each type of covariate are specified as follows: 
{\small
\begin{equation*} 
	\begin{aligned} 
		&\mathbf{X}_{1j} = \mathbf{U}_j, \quad &&1 \leq j \leq q_1; \\
		&\mathbb{P}(\mathbf{X}_{2_{ij}} = c \mid \mathbf{U}_{ij}) = \frac{\exp(w_c \mathbf{U}_{ij})}{\sum_{c^{'}=1}^{C} \exp(w_{c^{'}} \mathbf{U}_{ij})}, \quad &&c = 1,\ldots, C,\quad q_1 < j \leq q_1 + q_2; \\ 
		&\mathbf{X}_{3_{ij}} \mid \mathbf{U}_{ij}  \sim \mathrm{Poisson}(2 + 0.3,\mathbf{U}_{ij}), \quad &&q_1 + q_2 < j \leq q_1 + q_2 + q_3;\\ &\mathbb{P}(\mathbf{X}_{4_{ij}} = 1 \mid \mathbf{U}_{ij}) = \left\{1 + \exp(-0.5\mathbf{U}_{ij})\right\}^{-1}, \quad &&q_1 + q_2 + q_3 < j \leq p. 
	\end{aligned} 
\end{equation*}
}

To evenly distribute variable types across clients, we randomly shuffle the variable indices and partition the covariates into $K = 5$ groups, each of dimension $p_k = 20$, assuming a total number of covariates $p = 100$. The sparsity level is set as $s = 0.6$. For each client $k$, the true local regression coefficients are generated as $\beta^*_j = (-1)^j / 3$ for $1 \leq j \leq \lceil s p_k \rceil$, and $\beta^*_j = 0$ otherwise. The global regression coefficient vector is then constructed as $\bm{\beta}^* = (\bm{\beta}_1^{*\top}, \dots, \bm{\beta}_K^{*\top})^{\top}$.

For the response variable $Y$, we consider two regression settings. In the first setting, $Y$ is generated from a linear regression model as $Y_i = \mathbf{x}_i^{\top} \bm{\beta}^* + \epsilon_i$, where $\epsilon_i$ denotes independent random noise drawn from the standard normal distribution $\mathcal{N}(0, 1)$. In the second setting, $Y$ is generated from a logistic regression model, where the success probability is given by $\mathbb{P}(Y_i = 1) = {1 + \exp(-\mathbf{x}_i^{\top} \bm{\beta}^*)}^{-1}$.

\paragraph{Compared Methods}
To gain comparative insights, we consider three alternative methods. 
The first method, referred to as “VFL-ADMM-DP,” follows the approach proposed by \citet{hu2020learning}, where calibrated noise is injected into the local coefficient estimates during Algorithm~\ref{alg4} to perturb sensitive local linear embeddings and ensure privacy. The second method, termed “Vertigan” \citep{jiang2023distributed}, involves the generation of synthetic datasets on each client through a combination of local discriminators and a global generator; Algorithm~\ref{alg4} is then applied to the synthetic data for parameter estimation and variable selection. 
As performance benchmarks, we also include two additional settings based on unprotected data: (i) the “Original-C” scenario, where all data are stored and analyzed on a single client, and (ii) the “Original-VFL” scenario, where data remain distributed across clients without privacy mechanisms. These serve as upper-bound baselines for comparison.


\paragraph{Performance Measurements} The privacy budget takes values in $\{1,3,5\}$, each of which is evenly divided between $\varepsilon_1$ and $\varepsilon_2$.
For each simulation setting, a total of 100 replications are applied. 
In each replication,  the tuning parameter $\lambda$ is chosen by BIC criterion: $\text{BIC}(\lambda) = f(\widehat{\bm{\beta}}) + df_{\lambda}\cdot \log (N)$, where $f(\widehat{\bm{\beta}})$ is defined in Eq. \eqref{eq3}, and $df_{\lambda}$ is the
number of nonzero estimated coefficients in $\widehat{\bm{\beta}}$.
We consider three criteria to compare the performance of different methods. (1) Root Mean Square Error (RMSE) of the coefficient estimation, $\text{RMSE} = \sqrt{\|\hat{\bm{\beta}} - \bm{\beta}^*\|^2_2/p}$. (2) The geometric mean of sensitivity and specificity, which gives an overall variable selection performance measure, $\text{G-Means} = \sqrt{\text{SEN}\times \text{SPE}}$, where sensitivity measures the proportion of selected important covariates with $\text{SEN} = \sum_{j=1}^{p}\mathbb{I}(\hat{\beta}_j \neq 0, \beta^*_j \neq 0)/(\sum_{j=1}^{p}\mathbb{I}(\beta^*_j \neq 0))$, while specificity measures the proportion of nonselected unimportant covariates with $\text{SPE} = \sum_{j=1}^{p}\mathbb{I}(\hat{\beta}_j = 0, \beta^*_j = 0)/(\sum_{j=1}^{p}\mathbb{I}(\beta^*_j = 0))$. The value of $\text{G-Means}$ ranges from 0 to 1. This implies that a variable selection method works better if its value is closer to 1. (3) The false discovery rate (FDR), which represents the proportion of selected covariates that are actually irrelevant, with its definition being $\text{FDR} = \sum_{j=1}^{p}\mathbb{I}(\hat{\beta}_j \neq 0, \beta^*_j = 0)/\max (\sum_{j=1}^{p}\mathbb{I}(\hat{\beta}_j \neq 0), 1)$. We say that a method has better model selection performance if the FDR is smaller. 

\paragraph{Simulation Results}

Tables \ref{table1}-\ref{table3} report the RMSE of parameter estimates, the G-Means, and the FDR of variable selection for the competing methods under both linear and logistic regression in the VFL setting. Several conclusions can be drawn. First, for the Original-VFL approach, the RMSE and FDR converge to zero while the G-Means approaches one as the sample size increases, demonstrating that the proposed VFL-GLM-VS algorithm is consistent in both parameter estimation and support recovery, in accordance with Theorem \ref{vs-thm1}. Second, both the proposed VCDS and EVCDS methods substantially outperform the competing approaches. The poor performance of VFL-ADMM-DP indicates that simply injecting noise into standard optimization algorithms fails to achieve effective variable selection, while the inferior performance of the Vertigan method highlights the importance of exploiting interdependencies among variables across clients when constructing privatized data. Third, the estimation error and variable selection error of VCDS and EVCDS decrease as the sample size or privacy budget increases, approaching the results of Original-VFL. Nevertheless, a performance gap remains, reflecting the inevitable cost of privatization. Finally, VCDS and EVCDS generally perform comparably on complete datasets, with EVCDS showing a slight advantage, particularly in the logistic regression setting. This advantage can be attributed to its additional step of rearranging the sampled latent variables $\mathbf{Z}^{\text{new}}$ to preserve the rank information of the original data, thereby yielding privatized data of higher quality than that produced by VCDS.

\begin{table}[h]
	\centering
	\caption{The average RMSE of different methods for the linear and logistic regression coefficients based on 100 simulations, with standard error reported in parentheses.}
	\footnotesize
	\renewcommand\arraystretch{1.2}
	\tabcolsep=0.1cm
	\label{table1}
	\begin{tabular}{@{}cccccccc@{}}
		\Xhline{1px}
		$N$ & $\varepsilon$ & Original-C & Original-VFL   & VFL-ADMM-DP  & Vertigan  & VCDS & EVCDS      \\ 	\Xhline{1px}
		\multicolumn{8}{c}{Results of the linear model}    \\
		\multirow{3}{*}{$1000$} & $1$   & 0.0651 (0.0111) & 0.0714 (0.0133)  & 0.4383 (0.1551)  & 0.3241 (0.0685)    & 0.1490 (0.0098)
		  & 0.1487 (0.0169)             \\ 
		& $3$   & 0.0651 (0.0111) & 0.0714 (0.0133)   &  0.2366 (0.0371)  &  0.3172 (0.0732)    &  0.1374 (0.0117)  &  0.1391 (0.0171)     \\
		& $5$  & 0.0651 (0.0111) & 0.0714 (0.0133)    & 0.2111 (0.0171)  &  0.3153 (0.0705)  & 0.1244 (0.0112)  &  0.1331 (0.0602)       \\ \cline{2-8}
		\multirow{3}{*}{$3000$} & $1$ & 0.0355 (0.0073) & 0.0374 (0.0119) & 0.2794 (0.0565) & 0.3278 (0.0766) & 0.0871 (0.0158) & 0.0796 (0.0068)   \\
		& $3$ & 0.0355 (0.0073) & 0.0374 (0.0119) & 0.2141 (0.0102) & 0.3231 (0.0841) & 0.0707 (0.0098) & 0.0676 (0.0063) \\
		& $5$ & 0.0355 (0.0073) & 0.0374 (0.0119)  & 0.1848 (0.0048) & 0.3280 (0.0758) & 0.0741 (0.0152) & 0.0572 (0.0143) \\
		\cline{2-8}
		\multirow{3}{*}{$5000$} & $1$ & 0.0247 (0.0047) & 0.0246 (0.0065) & 0.2524 (0.0368) & 0.3206 (0.0704) & 0.0674 (0.0068)
		 & 0.0629 (0.0094) \\
		& $3$ &  0.0247 (0.0047) & 0.0246 (0.0065)  & 0.2134 (0.0069)  & 0.3182 (0.0752) & 0.0516 (0.0044) & 0.0547 (0.0224) \\
		& $5$ &  0.0247 (0.0047) & 0.0246 (0.0065) & 0.1924 (0.0034) & 0.3183 (0.0648) & 0.0522 (0.0075) & 0.0502 (0.0076) \\
		\Xhline{1px}
		\multicolumn{8}{c}{Results of the logistic model}    \\
		\multirow{3}{*}{$1000$} & $1$   & 0.1132 (0.0210) &	0.1163 (0.0172)  & 0.9639 (0.1247) &	0.3498 (0.3285 ) & 0.1888 (0.0040)	& 0.1639 (0.0096) \\
		& $3$ & 0.1132 (0.0210) &	0.1163 (0.0172)  & 0.3750 (0.0388) &	0.3482 (0.3279) & 0.1882 (0.0136) & 0.1467 (0.0219) \\
		& $5$ & 0.1132 (0.0210) &	0.1163 (0.0172)   &	0.2793 (0.0214) &	0.2913 (0.1714) &	0.1811 (0.0071)  & 0.1385 (0.0296) \\
		\cline{2-8}
		\multirow{3}{*}{$3000$} & $1$ & 0.0412 (0.0077) &	0.0511 (0.0086) & 0.4923 (0.0611) &	0.3276 (0.2209) & 0.1468 (0.0217) & 0.1315 (0.0204) \\
		& $3$ & 0.0412 (0.0077) &	0.0511 (0.0086) & 0.2570 (0.0173) &	0.3301 (0.3153) & 0.1425 (0.0143) & 0.1000 (0.0238) \\
		& $5$ & 0.0412 (0.0077) &	0.0511 (0.0086) & 0.2279 (0.0096) &	0.2784 (0.0961) &	0.1398 (0.0122)  & 0.0832 (0.0209)\\
		\cline{2-8}
		\multirow{3}{*}{$5000$} & $1$ & 0.0312 (0.0067) &	0.0391 (0.0091) & 	0.3772 (0.0383) &	0.2799 (0.0947) & 0.1413 (0.0187)
 & 0.1096 (0.0093) \\
		& $3$ & 0.0312 (0.0067) &	0.0391 (0.0091) & 0.2349 (0.0113) &	0.2779 (0.1083) & 0.1325 (0.0141) & 0.0908 (0.0119) \\
		& $5$ & 0.0312 (0.0067) &	0.0391 (0.0091) & 0.2195 (0.0066) &	0.2830 (0.1146) & 0.1328 (0.0172) & 0.0798 (0.0093) \\
		\Xhline{1px}
	\end{tabular}
\end{table}

\begin{table}[h]
	\centering
	\caption{The average G-Means of different methods for the linear regression coefficients based on 100 simulations, with standard error reported in parentheses.}
	\footnotesize
	\renewcommand\arraystretch{1.2}
	\tabcolsep=0.1cm
	\label{table2}
	\begin{tabular}{@{}cccccccc@{}}
		\Xhline{1px}
		$N$ & $\varepsilon$ & Original-C & Original-VFL   & VFL-ADMM-DP  & Vertigan  & VCDS & EVCDS      \\ 	\Xhline{1px}
		\multicolumn{8}{c}{Results of the linear model}    \\
		\multirow{3}{*}{$1000$} & $1$   & 0.9231 (0.0011) & 0.8740 (0.0012)  & 0.1761 (0.0061)  & 0.4939 (0.0028)    &  0.6938 (0.0019) & 0.7193  (0.0030)             \\ 
		& $3$   & 0.9231 (0.0011) & 0.8740 (0.0012)   &  0.3142 (0.0052)  &  0.4974  (0.0023)     &  0.7618 (0.0022) &  0.7992 (0.0020)       \\
		& $5$  & 0.9231 (0.0011) & 0.8740 (0.0012)    & 0.3924 (0.0030)  &  0.5022 (0.0026)      & 0.7817 (0.0027)  &  0.7959  (0.0483)     \\ \cline{2-8}
		\multirow{3}{*}{$3000$} & $1$ & 0.9712 (0.0004) & 0.9416 (0.0009) & 0.2575 (0.0043)  & 0.4842 (0.0023)  & 0.8911 (0.0005) &  0.9079  (0.0031) \\
		& $3$ & 0.9712 (0.0004) & 0.9416 (0.0009) & 0.4137 (0.0022) & 0.4958 (0.0027) & 0.9300 (0.0006) & 0.9546  (0.0003) \\
		& $5$ & 0.9712 (0.0004) & 0.9416 (0.0009)  & 0.4880 (0.0021) & 0.4899 (0.0023) & 0.9318 (0.0004) & 0.9661  (0.0004) \\
		\cline{2-8}
		\multirow{3}{*}{$5000$} & $1$ & 0.9800 (0.0002) & 0.9623 (0.0003) & 0.2961 (0.0031) & 0.4908 (0.0022) & 0.9463 (0.0006) & 0.9356 (0.0009) \\
		& $3$ &  0.9800 (0.0002) & 0.9623 (0.0003)  & 0.4581 (0.0026)  & 0.4947 (0.0023) &  0.9511 (0.0005) & 0.9621  (0.0010)  \\
		& $5$ &  0.9800 (0.0002) & 0.9623 (0.0003) & 0.5135 (0.0024) & 0.4952 (0.0020) & 0.9477 (0.0003)  & 0.9707   (0.0001) \\
		\Xhline{1px}
		\multicolumn{8}{c}{Results of the logistic model}    \\
		\multirow{3}{*}{$1000$} & $1$   & 0.8519 (0.0031) &	0.7594 (0.0042)  &  0.1193 (0.0047) &	0.1369 (0.0115) & 0.6909 (0.0199)  & 0.7218  (0.0061) \\
		 & $3$   & 0.8519 (0.0031) &	0.7594 (0.0042)  & 0.1981 (0.0055) &	0.1595 (0.0106) &	0.7023 (0.0123) & 0.7522  (0.0099) \\
		 & $5$   & 0.8519 (0.0031) &	0.7594 (0.0042)  & 0.2604 (0.0030) &	0.1341 (0.0110) &	0.7333 (0.0067) & 0.7563  (0.0120) \\
		 \cline{2-8}
		 \multirow{3}{*}{$3000$} & $1$ & 0.9703 (0.0003) &	0.9253 (0.0005) & 0.1616 (0.0047) &	0.1445 (0.0081) & 0.6909 (0.0199) & 0.8716 (0.0037) \\
		  & $3$ & 0.9703 (0.0003) &	0.9253 (0.0005) & 0.2938 (0.0032) &	0.1680 (0.0088) & 0.7333 (0.0073) & 0.9229 (0.0017) \\
		  & $5$ & 0.9703 (0.0003) &	0.9253 (0.0005) & 0.3647 (0.0029) &	0.1642 (0.0085) &	0.7023 (0.0218) & 0.9379  (0.0007) \\
		 \cline{2-8}
		 \multirow{3}{*}{$5000$} & $1$ & 0.9864 (0.0005) &	0.9516 (0.0003) & 0.2050 (0.0039) &	0.1618 (0.0088) & 0.8311 (0.0045) & 0.9430 (0.0006) \\
		  & $3$ & 0.9864 (0.0005) &	0.9516 (0.0003) & 0.3379 (0.0024) &	0.1567 (0.0106) & 0.8264 (0.0045) & 0.9414 (0.0006) \\
		  & $5$ & 0.9864 (0.0005) &	0.9516 (0.0003) & 0.4113 (0.0025) &	0.1709 (0.0087) & 0.8268 (0.0023) & 0.9466 (0.0007) \\
		 \Xhline{1px}
	\end{tabular}
\end{table}

\begin{table}[h]
	\centering
	\caption{The average FDR of different methods for the linear regression coefficients based on 100 simulations, with standard error reported in parentheses.}
	\footnotesize
	\renewcommand\arraystretch{1.2}
	\tabcolsep=0.1cm
	\label{table3}
	\begin{tabular}{@{}cccccccc@{}}
		\Xhline{1px}
		$N$ & $\varepsilon$ & Original-C & Original-VFL   & VFL-ADMM-DP  & Vertigan  & VCDS & EVCDS      \\ 	\Xhline{1px}
		\multicolumn{8}{c}{Results of the linear model}    \\
		\multirow{3}{*}{$1000$} & $1$   & 0.0115 (0.0002) & 0.0345 (0.0005)  & 0.4275 (0.0828) &  0.3997 (0.0029)   & 0.1651 (0.0014)  & 0.1716 (0.0053)    \\ 
		& $3$   & 0.0115 (0.0002) & 0.0345 (0.0005)   &  0.3622 (0.0279)  &  0.3952  (0.0024)    & 0.1018 (0.0009) &    0.0609 (0.0023)       \\
		& $5$  & 0.0115 (0.0002) & 0.0345 (0.0005)    &  0.1415 (0.0183) &  0.3885 (0.0028)     &  0.0849 (0.0024) &   0.1184 (0.0346)   \\ \cline{2-8}
		\multirow{3}{*}{$3000$} & $1$ & 0.0010 (0.0000) & 0.0089 (0.0002) & 0.3768 (0.0415) & 0.4105 (0.0026) & 0.0274 (0.0001) &  0.0325  (0.0006) \\
		& $3$ & 0.0010 (0.0000) & 0.0089 (0.0002) & 0.3817 (0.0077) & 0.3977 (0.0030) &  0.0254 (0.0001) & 0.0141 (0.0002) \\
		& $5$ & 0.0010 (0.0000) & 0.0089 (0.0002)  & 0.1622 (0.0044) & 0.4025 (0.0025) &  0.0250 (0.0004) & 0.0067  (0.0002) \\
		\cline{2-8}
		\multirow{3}{*}{$5000$} & $1$ & 0 (0) & 0.0006 (0.0000) & 0.3940 (0.0181) & 0.4007 (0.0024) & 0.0213 (0.0002) & 0.0215  (0.0000) \\
		& $3$ &  0 (0) & 0.0006 (0.0000)  & 0.3895 (0.0059) & 0.3975 (0.0026) & 0.0142 (0.0002) & 0.0108 (0.0001) \\
		& $5$ &  0 (0) & 0.0006 (0.0000) & 0.2068 (0.0031) & 0.3960 (0.0026) & 0.0142 (0.0002) & 0.0103 (0.0001) \\
		\Xhline{1px}
		\multicolumn{8}{c}{Results of the logistic model}    \\
		\multirow{3}{*}{$1000$} & $1$   & 0.0574 (0.0031) &	0.0330 (0.0021) & 0.4031 (0.1547) &	0.3998 (0.0000) & 0.3452 (0.0006) & 0.2097 (0.0036) \\
		& $3$   & 0.0574 (0.0031) &	0.0330 (0.0021) & 0.3901 (0.0671) &	0.3997 (0.0001) & 0.3188 (0.0085) & 0.1576 (0.0111) \\
		& $5$   & 0.0574 (0.0031) &	0.0330 (0.0021) & 0.3922 (0.0290) &	0.4006 (0.0000) & 0.3126 (0.0030) & 0.1523  (0.0135) \\
		\cline{2-8}
		\multirow{3}{*}{$3000$} & $1$ & 0.0012 (0.0000) &	0.0073 (0.0001) & 0.3713 (0.0951) &	0.4011 (0.0001) & 0.1114 (0.0125) & 0.0964 (0.0047) \\
		& $3$ & 0.0012 (0.0000) &	0.0073 (0.0001) & 0.3938 (0.0237) &	0.3993 (0.0001) & 0.1024 (0.0045) & 0.0532 (0.0020) \\
		& $5$ & 0.0012 (0.0000) &	0.0073 (0.0001) & 0.4003 (0.0137) &	0.4000 (0.0001) & 0.0923 (0.0103) & 0.0338 (0.0011) \\
		\cline{2-8}
		\multirow{3}{*}{$5000$} & $1$ & 0.0003 (0.0000) &	0.0050 (0.0001) & 0.3934 (0.0578) &	0.3991 (0.0001) & 0.0676 (0.0058) & 0.0107 (0.0002) \\
		& $3$ & 0.0003 (0.0000) &	0.0050 (0.0001) & 0.3930 (0.0141) &	0.4002 (0.0001) & 0.0310 (0.0046) & 0.0112  (0.0001) \\
		& $5$ & 0.0003 (0.0000) &	0.0050 (0.0001) & 0.4023 (0.0079) &	0.3999 (0.0001) & 0.0512 (0.0040) & 0.0111  (0.0001)\\
		\Xhline{1px}
	\end{tabular}
\end{table}

\subsection{Simulation Studies on Missing Dataset}\label{sec5-2}

We next investigate the performance of the proposed methods under client-wise missing data scenarios. Complete datasets are first generated following the procedure in Section~\ref{sec5-1}, and missingness is then introduced according to two mechanisms: MCAR and MAR, described below.

\textbf{\sc (1) Missing-completely-at-random (MCAR).}
Under MCAR, the missing probabilities for the five clients are set to $\rho_1=0.95,\; \rho_2=0.90,\; \rho_3=0.85,\; \rho_4=0.80,\; \rho_5=0.75$. For each sample $i$ and client $k$, the missing indicator $M_{ik}\in\{0,1\}$ (with $M_{ik}=1$ indicating that $\bm{x}_i^k$ is missing) is drawn independently as $M_{ik}\sim\mathrm{Bernoulli}(\rho_k)$ for $1\le i\le N,\;1\le k\le K$.

\textbf{\sc (2) missing-at-random (MAR).}
Under MAR, conditional on the observed data $(\bm{x}_{i,\mathrm{obs}},y_i)$, the missingness of $\bm{x}_{i,\mathrm{mis}}^k$ is independent of its own unobserved value but may depend on observed covariates and on the missingness indicators of other clients. To generate MAR patterns, we first draw a client-level mask
$\mathcal{M}=(M^1,\ldots,M^K)^\top$ with $M^k\sim\mathrm{Bernoulli}(0.5)$,
where $M^k=1$ indicates that the covariates on client $k$ may be missing and $M^k=0$ indicates they are always observed. Define
$\bm{\Delta}_{\mathrm{obs}}=\{k: M^k=0\}, \bm{\Delta}_{\mathrm{mis}}=\{k: M^k=1\}$.
For each client $k\in\bm{\Delta}_{\mathrm{mis}}$, we model the missingness probability by logistic regression. We consider two cases.

\vspace{2pt}
\noindent\textit{Simple case.} The missingness model is
$\operatorname{logit}\Pr(M_{ik}=1)
= 1 + \sum_{j\in\bm{\Delta}_{\mathrm{obs}}} (-1)^j\,\bm{\zeta}_j^\top\bm{x}_i^j,
\quad k\in\bm{\Delta}_{\mathrm{mis}}$,
where $\bm{\zeta}_j=(\zeta_{j1},\ldots,\zeta_{jp_j})^\top$ with $\zeta_{ja}=1/(j a)$ for $1\le a\le p_j$.

\vspace{2pt}
\noindent\textit{Complex case.} The missingness probability depends on observed covariates, the response, and the missingness of other clients:
$\operatorname{logit}\Pr(M_{ik}=1)
= 1 + \sum_{j\in\bm{\Delta}_{\mathrm{obs}}} (-1)^j\,\bm{\zeta}_j^\top\bm{x}_i^j
- y_i
+ \sum_{\substack{j'\in\bm{\Delta}_{\mathrm{mis}}\\ j'<k}} (-1)^{j'}\,M_{ij'},
\quad k\in\bm{\Delta}_{\mathrm{mis}}$.
Here the last term introduces dependence on prior clients' missingness (indexed by $j'<k$) to mimic more intricate cross-client missing patterns.

After datasets with missing values are generated, we apply VCDS/EVCDS/IEVCDS to obtain complete synthetic datasets, and then perform parameter estimation and variable selection using the VFL-GLM-VS algorithm. For comparison, we consider two baseline strategies: the \emph{Complete Case} (CC) method, which uses only samples with no missing values, and the \emph{Mean Imputation} (Impute) method, which fills missing entries with column means prior to analysis. For both baselines we also apply VCDS or EVCDS to produce privacy-protected versions, denoted CC-VCDS/CC-EVCDS and Impute-VCDS/Impute-EVCDS, respectively. For IEVCDS we set the number of iterations to $T=10$ to balance privacy cost and data utility. The privacy budget is set as $\varepsilon=1$.

\paragraph{Simulation Results}

Tables \ref{table4} and \ref{table5} report the performance of the competing methods based on the VCDS approach for the linear and logistic regression models under the two missing mechanisms, respectively. Tables \ref{table6} and \ref{table7} summarize the results for EVCDS and IEVCDS. Overall, the proposed methods generally achieve superior variable selection and estimation performance: they produce smaller RMSE and FDR and higher G-Means than the Complete Case (CC) and Mean Imputation (Impute) baselines. This advantage is especially pronounced under MAR, where the missing mechanism induces complex cross-client dependencies. The proposed approaches remain competitive even at high missing rates because they can better accommodate such dependencies and thus extract more information from incomplete samples. We also note that, under MCAR, the Impute method combined with VCDS or EVCDS attains comparable performance, which suggests that our privacy-preserving procedures can be applied on top of standard missing-data treatments without degrading the original method's effectiveness.

Comparing VCDS, EVCDS and IEVCDS yields several additional insights. First, VCDS performs very well for linear regression under MCAR: both selection and estimation errors decrease as sample size increases. However, VCDS deteriorates for logistic regression and under MAR, indicating its limitations when facing more complex response types or complicated missing mechanisms. Second, EVCDS produces results comparable to VCDS, with VCDS holding a slight advantage under MCAR. This is intuitive: the empirical CDF used by VCDS (based on original ranks) already yields consistent marginal estimates, whereas EVCDS relies on perturbed ranks and therefore incurs some loss in marginal estimation accuracy. Under MAR, however, EVCDS substantially outperforms VCDS in both model selection and coefficient estimation, which agrees with our theoretical analysis. Third, IEVCDS substantially improves upon EVCDS by iteratively exploiting the pseudo-complete data imputed during successive iterations, particularly under MAR. Under MCAR, iterations bring little improvement because the copula parameters are already consistently estimated from the original sample; under MAR, the initial copula estimates are biased, and iterative imputation combined with re-estimation of copula parameters can mitigate this bias, thereby improving the quality of the final privatized data and the performance of downstream inference.

\begin{table}[h]
	\centering
	\caption{The average results of VCDS methods for the linear regression model based on 100 simulations, with standard error reported in parentheses.}
	\footnotesize
	\renewcommand\arraystretch{1.2}
	\tabcolsep=0.1cm
	\label{table4}
	\begin{tabular}{@{}cccccccccc@{}}
		\Xhline{1px}
		\multirow{2}{*}{$N$} &  \multicolumn{3}{c}{CC-VCDS}  & \multicolumn{3}{c}{Impute-VCDS} &  \multicolumn{3}{c}{VCDS}      \\ 
				\cline{2-4} \cline{5-7} \cline{8-10} 
		 & RMSE & G-Means  & FDR      & RMSE & G-Means  & FDR     & RMSE & G-Means  & FDR             \\ 	\Xhline{1px}
		\multicolumn{10}{c}{MCAR}    \\
		\multirow{2}{*}{$1000$}   & 0.1815 & 0.5754 & 0.2424   &  0.1706 &	0.6620 & 0.1785  & 0.1536 &	0.6573 & 0.1745   \\ 
		& (0.0171) &  (0.0046) & (0.0042) &  (0.0135) &  (0.0039) &  (0.0028) &  (0.0136) &  (0.0017) &  (0.0011)  \\ 
		\multirow{2}{*}{$3000$}    & 0.1211 &	0.8121 & 0.0474 & 0.1196 &	0.8754 & 0.0561 & 0.0868 &	0.8901 &  0.0188  \\
		&  (0.0126) &  (0.0009) &  (0.0012) &  (0.0153) &  (0.0016)  & 	(0.0010)  & (0.0131) & 	(0.0004)  & (0.0010)  \\ 		
		\multirow{2}{*}{$5000$}  & 0.0965 &	0.8926 & 0.0156 & 0.0932 &	0.8902 & 0.0223 & 0.0727 &	0.9249  & 0.0151 \\
		& (0.0051) &  (0.0001) &  (0.0003)  &  (0.0138)  & 	(0.0004)  &  (0.0004)  & (0.0138) &  (0.0008)  &  (0.0003) \\ 
		\Xhline{1px}
		\multicolumn{10}{c}{MAR Simple}    \\
		\multirow{2}{*}{$1000$} & 0.2599 &	0.4897 & 0.2889 & 0.1701 &	0.5921 & 0.2234 & 0.1863 &	0.5555 & 0.2071 \\
		& (0.1466) &  (0.0084)  & (0.0232) &  (0.0077) &  (0.0359)  &  (0.0031) &  (0.0499)  & 	(0.0232) & 	(0.0120) \\ 
		\multirow{2}{*}{$3000$}& 0.1310 & 0.4640  & 0.0545 &  0.1385 &	0.7247 & 0.0613 & 0.1061 &	0.6881 & 0.0247 \\
		& (0.0334) &  (0.0250) & (0.0149) &  (0.0060) &  (0.0104) &  (0.0004) &  (0.0070) &  (0.0202) & (0.0008) \\ 
		\multirow{2}{*}{$5000$}& 0.1214 & 0.5353  & 0.0250 & 0.1367 & 0.7783 & 	0.0295  & 0.0781 &	0.8487 & 0.0228 \\
		& (0.0209) &  (0.0152) &  (0.0098) &  (0.0115) &  (0.0176) & (0.0032) &  (0.0075) &  (0.0004) &  (0.0004) \\ 
		\Xhline{1px}
		\multicolumn{10}{c}{MAR Complicated}    \\
		\multirow{2}{*}{$1000$} & 0.2310 &	0.3635 & 0.4238  & 0.1762 &	0.5402 & 0.2607 & 0.2100 &	0.4425 & 0.1381 \\ 
		& (0.0712) &  (0.0196) & (0.0525) &  (0.0111) & (0.0339) & 	(0.0163) &  (0.0994) & 	(0.0459)  &  (0.0219) \\ 		
		\multirow{2}{*}{$3000$} & 0.1829 &	0.5312 & 0.2213 & 0.1386 &	0.6600 & 0.1062 & 0.1227 &	0.6146 & 0.0617 \\
		& (0.0516)  &  (0.0184)  & 	(0.0152) & (0.0277) &  (0.0052) &  (0.0058) &  (0.0666)  & 	(0.0054) & 	(0.0019) \\ 
		\multirow{2}{*}{$5000$} & 0.1665 &	0.5499 &  0.0745 & 0.1250 &	0.6991 & 0.1057 & 0.0832 &	0.8202 & 0.0400 \\
		& (0.0428) &  (0.0275) &  (0.0195)  &  (0.0332)  & 	(0.0628)  &  (0.0079) &  (0.0253) & (0.0036) & 	(0.0020) \\ 
		\Xhline{1px}
	\end{tabular}
\end{table}

\begin{table}[h]
	\centering
	\caption{The average results of VCDS methods for the logistic regression model based on 100 simulations, with standard error reported in parentheses.}
	\footnotesize
	\renewcommand\arraystretch{1.2}
	\tabcolsep=0.1cm
	\label{table5}
	\begin{tabular}{@{}cccccccccc@{}}
		\Xhline{1px}
		\multirow{2}{*}{$N$} &  \multicolumn{3}{c}{CC-VCDS}  & \multicolumn{3}{c}{Impute-VCDS} &  \multicolumn{3}{c}{VCDS}      \\ 
		\cline{2-4} \cline{5-7} \cline{8-10} 
		& RMSE & G-Means  & FDR      & RMSE & G-Means  & FDR     & RMSE & G-Means  & FDR             \\ 	\Xhline{1px}
		\multicolumn{10}{c}{MCAR}    \\
		\multirow{2}{*}{$1000$}   &  0.2020  &	0.4832 & 0.3694 & 0.2026 &	0.4708 &  0.3458 & 0.1954 &	0.5456 & 0.3546   \\ 
		& (0.0072)  & 	(0.0046)  &  (0.0007) &  (0.0068)  &  (0.0407) &  (0.0037)  &  (0.0065)  &	(0.0042)  &	(0.0024)  \\ 
		\multirow{2}{*}{$3000$}    &  0.1722 &	0.6226 & 0.2839 & 0.1835 &	0.6157 & 0.1998 & 0.1751 & 0.7076 &  0.2374 \\
		&  (0.0224)  & 	(0.0172)  &  (0.0104)  &  (0.0159)  & 	(0.0192)  &  (0.0070) &  (0.0140) &  (0.0067) &  (0.0042)   \\ 		
		\multirow{2}{*}{$5000$}  & 0.1479 &	0.7639 & 0.1928 &  0.1438 &	0.7717 & 0.1877 & 0.1384 &	0.7944 &  0.1599  \\
		& (0.0087)  &  (0.0039)  & 	(0.0028)  &  (0.0119)  &  (0.0080)  &  (0.0059)  &  (0.0147)  &	(0.0067)  &  (0.0062) 	 \\ 
		\Xhline{1px}
		\multicolumn{10}{c}{MAR Simple}    \\
		\multirow{2}{*}{$1000$} & 0.2023 &	0.4963 & 0.3683 &  0.2037 &	0.5205 & 0.3664 &   0.1917 & 0.5673 &  0.3357  \\
		& (0.0045) &	(0.0017) &	(0.0006)  &  (0.0081) 	& (0.0359) &  (0.0031) &  (0.0092) & (0.0032)  &  (0.0023) 	 \\ 
		\multirow{2}{*}{$3000$}& 0.1777 & 0.6010 &	0.1432 & 0.1734 & 0.5957  & 0.2999  & 0.1579 &	0.6693  & 0.1726  \\
		&  (0.0137)  & 	(0.0118)  &	(0.0052)  &   (0.0228) &  (0.0185)  &  (0.0048)  &  (0.0167) &	(0.0080) &	(0.0036) 	\\ 
		\multirow{2}{*}{$5000$}& 0.1554 & 0.6542 	& 0.0442 &  0.1624 & 0.6824  &  0.1707  & 0.1426 &	0.7991 &  0.1666 \\
		& (0.0146) & (0.0164) &  (0.0023)  &  (0.0160) &  (0.0086) &  (0.0036)  & (0.0175) &  (0.0039) &  (0.0047)  \\ 
		\Xhline{1px}
		\multicolumn{10}{c}{MAR Complicated}    \\
		\multirow{2}{*}{$1000$} & 0.2052 &	0.5383 &	0.3218 & 0.2037 & 0.5356 &  0.3583 & 0.1973 & 0.5948 & 0.3184 \\ 
		& (0.0145)  &  (0.0044)  & 	(0.0048)  &  (0.0083) &  (0.0017)  &  (0.0006)   & (0.0138)  & 	(0.0006)  &  (0.0004) \\ 		
		\multirow{2}{*}{$3000$} & 0.1722 &	0.5314 & 0.1477 & 0.1888 &	0.5756 & 0.3143 & 0.1623 &	0.6970 &	0.1852  \\
		& (0.0080) &  (0.0040)  &	(0.0020) &  (0.0130) &	(0.0099) & 	(0.0024) & (0.0236) &  (0.0107)  & 	(0.0055)  \\ 
		\multirow{2}{*}{$5000$} & 0.1593 &	0.7602 	& 0.1680 &  0.1773 & 0.6857  & 0.2531 & 0.1394 & 0.8171 &  0.1495 	 \\
		& (0.0169) &  (0.0030)  & (0.0041)  & (0.0143) &  (0.0092) &  (0.0036) &  (0.0186) &  (0.0053) &  (0.0068)  \\ 
		\Xhline{1px}
	\end{tabular}
\end{table}

\begin{sidewaystable}
	\centering
	\footnotesize
	\caption{The average results of EVCDS and IEVCDS methods for the linear regression model based on 100 simulations, with standard error reported in parentheses.}
	\label{table6}
	\renewcommand\arraystretch{1.2}
	\begin{tabular}{@{}c ccc ccc ccc ccc@{}}
		\Xhline{1px}
		\multirow{2}{*}{\makecell[c]{$N$}} & \multicolumn{3}{c}{CC-EVCDS}        & \multicolumn{3}{c}{Impute-EVCDS}    & \multicolumn{3}{c}{EVCDS}             & \multicolumn{3}{c}{IEVCDS}       \\
		\cline{2-13}
		& RMSE & G-Means & FDR & RMSE & G-Means & FDR & RMSE & G-Means & FDR & RMSE & G-Means & FDR \\ \Xhline{1px}
		\multicolumn{13}{c}{MCAR (full obs. = 77.32\%)}  \\
		\multirow{2}{*}{$1000$}   & 0.2009 & 0.5659  & 0.3181 & 0.1682 & 0.6771 & 0.2127 & 0.1673 &	0.6928 & 0.2179 & 0.1907 &	0.6908 	& 0.2178 \\
		& (0.0069)  & 	(0.0031)  &  (0.0004) &  (0.0192) &  (0.0010)  &  (0.0015)  & (0.0148)  &  (0.0023)  & 	(0.0064) & (0.0135) 	& (0.0037)  & (0.0034)  \\ 
		\multirow{2}{*}{$3000$}   &   0.1455 &	0.7501  &	0.1533  &  0.1146 &	0.8814 &	0.0308 & 0.0987 &	0.8612 &	0.0230 &
		0.1823 	& 0.8216 &	0.1323 \\
		& (0.0294) &  (0.0096) &  (0.0111) & (0.0188)  &  (0.0040) & (0.0007) & (0.0220) & 	(0.0055) & 	(0.0010) & (0.0159) & (0.0038) & 	(0.0045) \\ 	
		\multirow{2}{*}{$5000$}   &  0.1142 &	0.8764 &	0.0388 &  0.1064 &	0.9265 & 	0.0254 & 0.0904 &	0.8510 &	0.0590 & 0.1675 &	0.8775 & 	0.0652 \\
		& (0.0152) &  (0.0013) & (0.0005) & (0.0218)  & (0.0005)  & (0.0001) & (0.0329) &  (0.0249)  & 	(0.0115) & (0.0020) &  (0.0049) &  (0.0029)   \\ \Xhline{1px}
		\multicolumn{13}{c}{MAR Simple (full obs. = 44.38\%)}       \\
		\multirow{2}{*}{$1000$}    &  0.2135 &	0.5631 &	0.3040 & 0.1669 &	0.6788  &	0.2054 & 0.1905 &	0.6451 & 	0.2544 & 0.1831 &	0.6911 &	0.2196 \\
		& (0.0218)  & 	(0.0051)  &  (0.0065)  &  (0.0104) &  (0.0031) &  (0.0018) & (0.0168) &	(0.0249) & 	(0.0239) & (0.0114) &	(0.0025) &	(0.0025) \\ 
		\multirow{2}{*}{$3000$}   &  0.1515 &	0.7331 &	0.1526 & 0.1553 &	0.8037 &	0.1554 & 0.1479 &	0.7925 &	0.1210 & 0.1511 &	0.8343 & 	0.0854 \\
		& (0.0325) &  (0.0087)  & (0.0103)  &  (0.0122)  & 	(0.0023)  &  (0.0013) & (0.0539)  &  (0.0053)  &  (0.0038)  & (0.0296)  	& (0.0027)  & (0.0055)  	\\
		\multirow{2}{*}{$5000$}   & 0.1376 &	0.7892 &	0.0964 & 0.1389 &	0.8531 &	0.0895 & 0.1240 &	0.8743 &	0.0327 & 0.1355 &	0.9026 &	0.0587 \\
		& (0.0501)  &  (0.0318)  & 	(0.0243)  &  (0.0241)  & (0.0033)  &  (0.0084)  & (0.0248)  & (0.0014)  & (0.0005)  & (0.0107) &  	(0.0001)  &  (0.0005)  \\ \Xhline{1px}
		\multicolumn{13}{c}{MAR Complicated (full obs. = 23.79\%)}      \\
		\multirow{2}{*}{$1000$}  & 0.2723 &	0.5006 &	0.3612 & 0.1719 &	0.6705 &	0.2428 & 0.1787 &	0.6167 &	0.2362 & 
		0.1843 &	0.6854 &	0.1759 \\
		& (0.0606)  &  (0.0085)  & 	(0.0073) &  (0.0036)  &  (0.0004)  &  (0.0004)  & (0.0194) &  (0.0060) & (0.0076) & (0.0222) & 	(0.0049)  &  (0.0054) \\ 
		\multirow{2}{*}{$3000$}  & 0.1770 &	0.6914 & 	0.2196 & 0.1477 &	0.7944 &	0.1651 & 0.1340 &	0.8386 &	0.0910 & 0.1318 	& 0.8855 	& 0.0357 \\
		& (0.0224) &  (0.0091)  &  (0.0112)  & (0.0178)   & 	(0.0045)  &  (0.0055) & (0.0358)  &	(0.0018)  &  (0.0010) & (0.0268) & 	(0.0008)  &  (0.0009)  \\
		\multirow{2}{*}{$5000$}  & 0.1542 &	0.7421 &	0.1633  & 0.1518 &	0.7897 &	0.1711 & 0.1156 &	0.8857 	& 0.0487 & 0.1332 &	0.9468 &	0.0096 \\
		& (0.0353)  & 	(0.0079)  &  (0.0100)  & (0.0322)  &  (0.0112)  & 	(0.0096)  & (0.0308)  & (0.0015)  &  (0.0010)  & (0.0024) &  (0.0055)  & 	(0.0041) 
	  \\  \Xhline{1px}
	\end{tabular}
\end{sidewaystable}

\begin{sidewaystable}
	\centering
	\footnotesize
	\caption{The average results of EVCDS and IEVCDS methods for the logistic regression model based on 100 simulations, with standard error reported in parentheses.}
	\label{table7}
	\renewcommand\arraystretch{1.2}
	\begin{tabular}{@{}c ccc ccc ccc ccc@{}}
		\Xhline{1px}
		\multirow{2}{*}{\makecell[c]{$N$}} & \multicolumn{3}{c}{CC-EVCDS}        & \multicolumn{3}{c}{Impute-EVCDS}    & \multicolumn{3}{c}{EVCDS}             & \multicolumn{3}{c}{IEVCDS}       \\
		\cline{2-13}
		& RMSE & G-Means & FDR & RMSE & G-Means & FDR & RMSE & G-Means & FDR & RMSE & G-Means & FDR \\ \Xhline{1px}
		\multicolumn{13}{c}{MCAR (full obs. = 77.32\%)}  \\
		\multirow{2}{*}{$1000$}   & 0.2041 &	0.5694 &	0.3206 & 0.1927 &	0.5605 &	0.2959 & 0.1785 &	0.6351 &	0.2823 & 0.1942 &	0.6833 &	0.2489  \\
		& (0.0155) & 	(0.0029)  & 	(0.0040)  & (0.0099)  &  (0.0029)  &  (0.0022)  & (0.0058)  & 	(0.0020)  & 	(0.0015) & (0.0061)  & 	(0.0030)  & 	(0.0021)  \\ 
		\multirow{2}{*}{$3000$}   &  0.1620 &	0.7722 & 	0.1633 & 0.1561 &	0.8230 &	0.1442 & 0.1238 &	0.8529 &	0.1090 & 0.1677 	& 0.8500 &	0.1053   \\
		& (0.0080) &  (0.0054)  & 	(0.0056)  & (0.0066)  &  (0.0015)  &  (0.0015)  & (0.0191)  & 	(0.0023)  &  (0.0019) &  (0.0121) & 	(0.0040)  & 	(0.0032) \\ 	
		\multirow{2}{*}{$5000$}   &  0.1372 &	0.8377 &	0.0928 & 0.1322 &	0.9268 &	0.0322 & 0.1106 &	0.8942 &	0.0419 & 0.1650 	& 0.8746 &	0.0727  \\
		&   (0.0073)  &  (0.0006)  &  (0.0009)  & (0.0064)  & 	(0.0005)  &  (0.0001)  & (0.0067)  &  (0.0012)  & 	(0.0003) & (0.0108)  &  (0.0036)  &  (0.0023)  \\ \Xhline{1px}
		\multicolumn{13}{c}{MAR Simple (full obs. = 44.38\%)}       \\
		\multirow{2}{*}{$1000$}    & 0.2139 &	0.6019 &	0.2863 & 0.1848 &	0.6111 &	0.3070 & 0.1759 &	0.6145 &	0.2980 & 0.1635 &	0.6693 &	0.2513   \\
		& (0.0262)  & 	(0.0014)  &  (0.0014)  & (0.0107)  &  (0.0006)  &  (0.0003)  & (0.0065)  & 	(0.0022)  &  (0.0022) & (0.0014) &  	(0.0014)  & 	(0.0009)  \\ 
		\multirow{2}{*}{$3000$}   &  0.1474 &	0.7585  &	0.1805 & 0.1605 &	0.7861 &	0.1806 & 0.1311 &	0.8095 &	0.1448 & 0.1365 &	0.8545  &	0.0937 	 \\
		&   (0.0166)  & 	(0.0027)  & 	(0.0036)  & (0.0172)  & 	(0.0046)  & 	(0.0035) & (0.0175)  & 	(0.0042)  & 	(0.0046) & (0.0121)  & 	(0.0006)  & 	(0.0011) 	\\
		\multirow{2}{*}{$5000$}   & 0.1193 &	0.8406 &	0.0757 & 0.1457 &	0.8540 &	0.1186 & 0.1130 &	0.8657 &	0.0751 & 0.1165 &	0.9022 &	0.0413  \\
		&  (0.0185)  & 	(0.0011)  & 	(0.0022)  & (0.0108)  & 	(0.0021)  & 	(0.0014)  & (0.0106)  &  (0.0024)  &  (0.0008) & (0.0171)  & 	(0.0012)  & 	(0.0014) 	 \\ \Xhline{1px}
		\multicolumn{13}{c}{MAR Complicated (full obs. = 23.79\%)}      \\
		\multirow{2}{*}{$1000$}  & 0.2490 &	0.5345 &	0.3614 & 0.1819 &	0.6082 &	0.3102 & 0.1842 &	0.6179 &	0.2944 & 0.1776 	& 0.6659 	& 0.2534  \\
		& (0.0496)  & 	(0.0107)  &  (0.0118)  & (0.0097)  &  (0.0047) &  (0.0031)  & (0.0056)  & 	(0.0056)  &  (0.0043) & (0.0060) &  	(0.0021)  & 	(0.0021) 	 \\ 
		\multirow{2}{*}{$3000$}  & 0.1415 &	0.7600 &	0.1499 & 0.1634 &	0.7684 &	0.1927 & 0.1275 &	0.8098 &	0.1485 & 0.1291 &	0.8665 &	0.0909  \\
		&  (0.0137)  & 	(0.0039) & 	(0.0026)  & (0.0178)  & 	(0.0111)  &  (0.0059)  & (0.0261)  & 	(0.0049)  & 	(0.0053) & (0.0134)  & 	(0.0002)  & (0.0019) 	 \\
		\multirow{2}{*}{$5000$}  &  0.1281 	& 0.7964 &	0.0894 & 0.1470 &	0.8444 &	0.1285 & 0.1082 &	0.8612 &	0.0632 & 0.1185 	& 0.9025 &	0.0553 	\\
		& (0.0101)  & 	(0.0010)  &  (0.0004) & (0.0106)  &  (0.0040)  & 	(0.0028)  & (0.0156)  & 	(0.0011)  & 	(0.0010) & (0.0136)  & 	(0.0005)  & 	(0.0009) 
	\\  \Xhline{1px}
	\end{tabular}
\end{sidewaystable}

\section{Real Data Application}
In this section, we apply the proposed method to a case study on regulatory risk assessment for Small and Medium-sized Enterprises (SMEs). SMEs are businesses whose employee numbers, revenues, or assets fall below specific thresholds defined by national or regional standards. They account for a significant proportion of all enterprises in industrialized economies and contribute substantially to job creation and economic resilience.

Compared to larger firms, SMEs tend to exhibit distinct operational characteristics. Notably, they often show lower compliance rates with health, safety, and environmental regulations. Effectively regulating SMEs presents a substantial policy challenge for regulatory agencies, primarily due to several inherent constraints that limit the applicability of conventional regulatory approaches. These include a general lack of environmental awareness and technical expertise, low public visibility, and their sheer volume, which results in infrequent inspections and leaves many SMEs beyond the reach of policy initiatives.
Given the limited resources of most regulatory bodies, it is crucial to design strategies that can bring the majority of SMEs into regulatory compliance in an efficient and scalable manner.

A primary goal of this study is to assess the regulatory risk faced by SMEs and to identify key factors that influence such risk.
To achieve this, we integrate five complementary sources of SME-related information, which were originally maintained in separate departments or institutions. These data sources include:
(1) financial records from the Credit Agency;
(2) annual inspection data from the Market Supervision and Administration Bureau;
(3) records of business anomalies from the Judicial Authority;
(4) registration details from the Bureau of Industry and Commerce; and
(5) administrative penalty records from the Law Enforcement Agency.
This integration yields a dataset comprising 55 variables for 71,827 SMEs across various industries.
A summary of the SMEs dataset is provided in Table \ref{table-d}, and a detailed description of all variables is available in Appendix of the supplementary materials.

\begin{table}[h]
	\centering
	\caption{Summaries of the SMEs dataset.}
	\footnotesize
	\tabcolsep=0.1cm
	\label{table-d}
	\begin{tabular}{@{}cccc@{}}
		\toprule
		Client index & Source of information & Number of variables & Missing proportion \\ \midrule
		1 & Credit Agency & 13 & 72.34\% \\
		2 & Market Supervision and Administration Bureau & 6 & 83.53\% \\
		3 & Judicial Authority & 6 & 87.33\% \\
		4 & Bureau of Industry and Commerce & 19 & 27.11\% \\
		5 & Law Enforcement Agency & 11 & 0.00\% \\
		\bottomrule
	\end{tabular}
\end{table}

Since the dataset is compiled from five distinct institutions, it is not feasible to centralize the data for traditional analysis due to data security and privacy concerns. Therefore, we adopt the VFL framework, in which each institution is treated as an independent client. A notable feature of the dataset is the presence of client-wise missingness across nearly all clients, with missing rates ranging from 27.11\% to 87.33\%. This type of missingness arises due to the high cost of data acquisition or institutional limitations. For example, the sheer number of SMEs vastly exceeds the capacity of inspectors, making it impractical to conduct comprehensive inspections. As a result, many SMEs lack records in the Market Supervision and Administration Bureau. Consequently, only 184 enterprises in the dataset have complete covariate information across all five institutions.

In our analysis, we treat the variable ``whether an enterprise receives administrative penalties by the end of the year", provided by the fifth client, as the binary response variable $Y \in \{0,1\}$. Our objective is to identify important covariates associated with $Y$ that may help predict future non-compliance behavior among SMEs. To this end, we construct a vertically federated logistic regression model under the sparsity assumption to analyze regulatory risk.

\begin{figure}[h]
	\centering
	\includegraphics[width=0.7\textwidth]{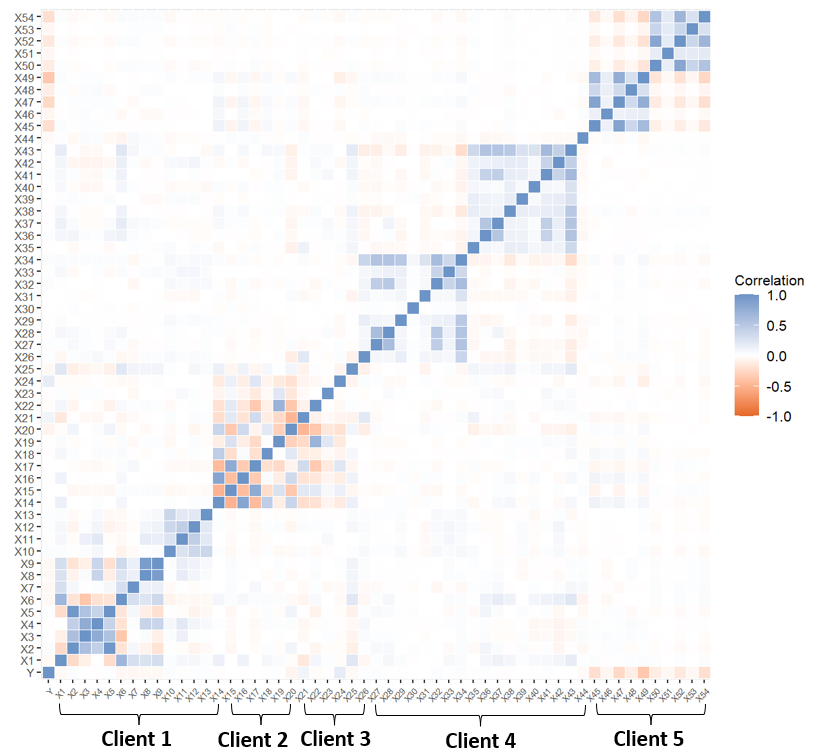} \hfill
	\caption{The rank correlation map between covariates across five clients. }
	\label{f-corr}
\end{figure}

Before building the model, we conduct a descriptive analysis of the data. 
Due to the presence of missing values, we compute the marginal Kendall's rank correlation between the response and each covariate using all available pairs of observations. Figure~\ref{f-corr} illustrates the correlation structure among covariates across different clients. While covariates within the same client generally have stronger correlations, there remains substantial cross-client dependency, which we aim to exploit to improve both prediction accuracy and variable selection performance.

We consider three competing methods. The first is the ``Single'' method, in which the analysis is conducted solely on the fifth client, where the response variable resides. The second is the ``CC'' (Complete Case) method, which uses only the samples that are fully observed across all institutions for vertically federated learning. The third is the ``Impute'' method, where missing values are imputed using the mean of the observed entries, and the resulting dataset is then used for downstream analysis.
To enhance data privacy, we apply the proposed VCDS/EVCDS approach to each method to generate privatized datasets. To align with the simulation study, the privacy budget is fixed at 1 and the iteration step at 10.
The SCAD regularizer is adopted to induce sparsity in the estimated coefficients.
For performance comparison, we randomly split the data into training and test sets 10 times. In each split, the test set consists of 92 randomly selected complete observations, while the training set includes the remaining complete cases along with samples that contain missing covariates.

\begin{table}[h]
	\centering
	\caption{Average prediction and variable selection results for the vertically federated logistic regression model based on 10 simulations. Standard errors are reported in parentheses.}
	\footnotesize
	\renewcommand\arraystretch{1.2}
	\tabcolsep=0.1cm
	\label{table8}
	\begin{tabular}{@{}ccccccccccc@{}}
		\Xhline{1px}
		\multirow{2}{*}{Method} &  \multicolumn{4}{c}{VCDS}  & &\multicolumn{4}{c}{EVCDS}  & \multirow{2}{*}{IEVCDS}    \\ 
		\cline{2-5} \cline{7-10}  
		& Single & CC  & Impute      & VCDS &   & Single     & CC  & Impute      & EVCDS &         \\ 	\Xhline{1px}
		\multirow{2}{*}{Recall}   & 0.0194  & 0.2761  & 0.6348  &  0.8190   &   & 0.5772 & 0.5945 &  0.7857  &  0.9113 & 0.9533   \\ 
		& (0.0254) & (0.1482) & (0.0825) & (0.0769) &  & (0.0652) & (0.1722) & (0.0564) & (0.0288) & (0.0493) \\
		\multirow{2}{*}{AUC}    & 0.5446  & 0.5202   &  0.5414  &  0.6607    &   & 0.5304  & 0.5534  & 0.5837  &  0.7156 & 0.7173     \\
		&  (0.0461) & (0.0259)    &  (0.0442) &  (0.0503)     &   &  (0.0360) & (0.0737) &  (0.0502) & (0.0227) & (0.0247)  \\ 
		NS  & 8  & 8 & 33 & 17 &  & 6 & 7 & 27 & 14 & 13 \\
		\Xhline{1px}
	\end{tabular}
 \vspace{0.5em}
\parbox{0.8\linewidth}{
	\footnotesize \textit{Note:} The ``NS'' represents the average number of selected variables.
}
\end{table}

For the assessment of model prediction performance, we calculate Recall and Area Under the Receiver Operating Characteristic Curve (AUC) for each test set corresponding to each method. The Recall is defined as $\text{Recall} = \text{TP}/(\text{TP} + \text{FN})$, where $\text{TP}$ denotes the number of risky SMEs correctly predicted as risky, and $\text{FN}$ is the number of risky SMEs wrongly predicted as compliant. 
Recall measures the model's ability to identify all truly positive instances. In the context of predicting regulatory violations among SMEs, a higher recall indicates that the model can successfully detect the majority of potentially non-compliant enterprises. This is particularly important in the early-stage risk screening phase, where large-scale identification of high-risk entities is prioritized, typically followed by manual review and further investigation.
In addition, we report the AUC, a standard evaluation metric that quantifies the model's overall ability to discriminate between risky and compliant enterprises across all possible classification thresholds. 

 Table \ref{table8} reports the performance of all methods based on 10 replications. The competing methods shown in the left panel of Table \ref{table8} are incorporated into VCDS, while those in the right panel are incorporated into EVCDS. Overall, the proposed methods achieve higher Recall and AUC values with substantially smaller model sizes, suggesting that they deliver superior variable selection and prediction accuracy compared with all other methods for the SMEs data, owing to their ability to incorporate correlation information from incomplete observations. Comparing VCDS, EVCDS, and IEVCDS, we observe that EVCDS outperforms VCDS, indicating that the missingness mechanism is more complex than MCAR, and that the MCAR-based VCDS method cannot fully address such complex missing patterns. Furthermore, IEVCDS further improves upon EVCDS, which is consistent with the theoretical analysis and simulation results. The CC method consistently selects 7–8 variables, as there are only 92 complete observations in the training sets. In contrast, the Impute method selects considerably more variables than all other methods.
 
 Table 2 in the supplementary materials lists the top NS variables most frequently selected by each method across the 10 training sets, where NS denotes the average number of variables selected by the corresponding method. The 13 variables selected by the IEVCDS method include: financial variables such as ROE (return on equity) and Scale from the Credit Agency; FOI-3y (number of inspections received in the past three years) and CI-3y (number of compliant inspections in the past three years) from the Market Supervision and Administration Bureau; ABNAP (frequency of not submitting annual reports) and Removed (frequency of being removed from the abnormal business record) from the Judicial Authority; SCOPE-1y and SCOPE-3y (frequency of short- and long-term changes in business scope), and Industry of SMEs from the Bureau of Industry and Commerce; as well as Punished (whether administrative penalties were received), FP2-3y (frequency of mild penalties in the past three years), FP3-3y (frequency of severe penalties in the past three years), and FAP (frequency of administrative penalties) from the Law Enforcement Agency. Most of these variables are also selected by EVCDS and VCDS. These covariates are important indicators of SMEs' compliance behavior, as confirmed in prior studies \citep{mohamad2014does, valaskova2018financial, zhou2021does}. Notably, the proposed methods can select variables from clients with high missing rates, whereas the CC method tends to select variables with more complete information, mostly from clients 1, 4, and 5, whose missing rates are relatively lower. By contrast, the Impute method selects many redundant variables.
 
 In summary, the proposed methods achieve higher Recall and AUC in test sets than other competing approaches, while selecting fewer variables. This demonstrates their advantage in both variable selection and predictive performance. Moreover, the selected covariates are indeed relevant to the response variable, which is further supported by findings in economics and management research.

\section{Conclusion}

Vertical Federated Learning has emerged as a promising paradigm for collaborative learning with distributed data. 
In practice, data across clients in VFL often suffer from client-wise missingness, where entire covariate blocks held by certain clients are unobserved for some samples, resulting in a large proportion of unaligned data. 
Meanwhile, conventional modeling and analysis under the VFL framework are prone to privacy leakage risks. 
To fully exploit unaligned samples while safeguarding sensitive information, it is essential to incorporate data privatization under the missingness constraint. 
In this work, we develop a VFL data privatization framework that addresses client-wise missingness using a Gaussian copula model. 

We first introduce a private estimation procedure for the copula correlation structure and the marginal distributions. 
Building on these components, we address data privatization under both MCAR and MAR mechanisms. 
Specifically, under MCAR, the correlation matrix and ECDFs privately estimated from the observed data are consistent. 
In this setting, complete privatized data can be generated by resampling Gaussian scores and transforming them back to the original data domain, a method we term VCDS. 
However, VCDS is not applicable under MAR. 
To tackle this challenge, we propose a nonparametric marginal estimation technique that yields consistent CDF estimates even under MAR. 
This leads to the EVCDS algorithm, which generates privatized data based on observed samples. 
Nevertheless, EVCDS still suffers from bias caused by MAR, as the initial correlation matrix estimate remains biased. 
To address this issue, we further introduce the IEVCDS algorithm, which iteratively refines the copula parameter estimates using pseudo-complete data, and generates the final privatized dataset based on the updated parameters. 
We then investigate the utility of the privatized data for GLM coefficient estimation and variable selection under VFL. 

From a theoretical perspective, we establish the relationship between privacy guarantees and estimation errors for both the correlation matrix and marginal distributions in the copula model. 
We formally define Vertical Distributed Attribute Differential Privacy (VDADP) for VFL and show that, under this definition, the proposed methods provide rigorous differential privacy guarantees as well as utility guarantees measured by the KL divergence between the original and privatized distributions. 
Furthermore, we prove asymptotic normality, estimation consistency, and variable selection consistency for VFL-GLMs fitted to the privatized data. 
The effectiveness of our methods is demonstrated through extensive simulations and real data analysis.

\bibliographystyle{asa}
\bibliography{reference}

\end{document}